\documentclass[preprint,12pt]{elsarticle}

\usepackage{amsfonts,amssymb,amsmath}
\usepackage[dvipsnames]{xcolor}
\usepackage{ulem}
\usepackage{comment}
\usepackage{hyperref}

\usepackage{amsmath}
\numberwithin{equation}{section}

\newcommand{\SC}[1]{[{\color{blue}#1}]}

\definecolor{mathblue}{rgb}{0.368417, 0.506779, 0.709798}
\definecolor{mathorange}{rgb}{0.880722, 0.611041, 0.142051}
\definecolor{mathgreen}{rgb}{0.560181, 0.691569, 0.194885}
\definecolor{mathred}{rgb}{0.922526, 0.385626, 0.209179}
\definecolor{mathviolet}{rgb}{0.528488, 0.470624, 0.701351}
\definecolor{mathbrown}{rgb}{0.772079, 0.431554, 0.102387}

\def \BE {\begin{equation}}
\def \EE {\end{equation}}
\def \BEA {\begin{eqnarray}}
\def \EEA {\end{eqnarray}}

\def \one {^{(1)}}
\def \two {^{(2)}}

\def \e {{\epsilon}}
\def\be{\begin{equation}}
\def\ee{\end{equation}}

\journal{Physics Reports}

\begin{document}

\title{Wave Turbulence and thermalization in one-dimensional chains}

\author{M. Onorato$^{1,2}$, Y. V. Lvov $^3$, G. Dematteis$^{4,3}$, 
and S. Chibbaro$^{5}$\\
$^1$  Dipartimento di Fisica, Universit{\`a} degli Studi di Torino, Via P. Giuria 1, 10125 Torino, Italy
\\
$^2$ INFN, Sezione di Torino, Via Pietro Giuria 1, 10125 Torino, Italy
\\
$^3$Department of Mathematical Sciences, Rensselaer Polytechnic Institute, Troy, New York 12180, USA
\\
$^4$Physical Oceanography Department, Woods Hole Oceanographic Institution, Woods Hole, Massachusetts 02453, USA
\\
$^5$ Universit\'e Paris-Saclay, CNRS, UMR 9015, LISN, F - 91405 Orsay cedex, France }


\begin{abstract}
One-dimensional chains are used as a fundamental model of condensed matter, and have constituted the starting point for key developments in nonlinear physics and complex systems.
The pioneering  work in this field was proposed by Fermi, Pasta, Ulam and Tsingou in the 50s in Los Alamos.
An intense and fruitful  mathematical and physical research followed during these last 70 years. Recently, a fresh look at the mechanisms at the route of thermalization of such systems has been provided  through the lens of the Wave Turbulence approach. 
In this review, we give a critical summary of the results obtained in this framework.
We also present a series of open problems and challenges that future work needs to address.

\end{abstract}

\maketitle


\begin{highlights}
\item Research highlight 1
\item Research highlight 2
\end{highlights}

\begin{keyword}



\end{keyword}

\tableofcontents
\vspace{2cm}




\section{Introduction}
\label{sec:1}

In the early fifties in Los Alamos, Enrico Fermi, John Pasta,
Stanislav Ulam and Mary Tsingou studied numerically a
one dimensional monoatomic  lattice, including cubic and quartic
anharmonic potentials~\cite{fermi1955alamos}.  This prototypical
one-dimensional chain was thought by Fermi and collaborators to be a
very interesting model to be investigated using the at-that-time new
computer MANIAC-I.  They proposed this problem in relation to the
foundations of statistical mechanics and the ergodic
problem~\cite{poincare1893methodes,ma1985statistical,castiglione2008chaos,chibbaro2014reductionism},
an issue that Fermi had tried to address at the beginning of his
career~\cite{fermi1923dimostrazione}.  Yet, as witnessed by Ulam,
Fermi was firmly convinced that, despite its simplicity and high
degree of idealisation, such one-dimensional chains should represent a
fertile tool to investigate many problems in statistical mechanics and
solid state physics~\cite{gallavotti2007fermi,benenti2020anomalous}.

At that time, it was understood that linear systems, characterized by
an harmonic potential, cannot reach a thermalized state. The
presence of a small nonlinearity was expected to lead the system to
thermalization, in line with previous mathematical results by
Poincar\'e and Fermi himself.  The project was started by Fermi,
Pasta, Ulam, and, with the contribution of Tsingou,
a code for solving the equations of motion was set up and the output of 
the simulations was graphed (the role of Mary Tsingou has 
been highlighted in \cite{dauxois2008fermi}, and, nowadays, it is customary to 
use the initials FPUT, instead of FPU).
Contrary to expectations, thermalization
among the modes was not observed.  Here, we quote the last
sentence of the abstract of~\cite{fermi1955studies}: ``The results show
very little, if any, tendency toward equipartition of energy among the
degrees of freedom''.
They observed that the initially perturbed mode was sharing energy
with other modes but, after some times, the energy was, almost fully,
transferred back to the initial mode: this phenomenon is now known as the
Fermi-Pasta-Ulam-Tsingou recurrence.  Fermi himself showed enthusiasm
for the finding, manifesting the conviction that a ``little'' discovery
was made, see the preface of Ulam in \cite{fermi1955studies}.  Unfortunately, Fermi died  shortly after. The
results of the work found place in an internal report whose preprint
was shared with colleagues, but was not published, and remained rather unnoticed. 
Ulam presented
  the work at some conferences, and only 10 years later the report
  was printed in the collection of Fermi's works \cite{persico1967collected}.

As a matter of fact, the FPUT experiment has been widely influential. First of all, it was one of the very first important numerical experiments in physics, and 
basically launched the use of numerics for solving the dynamical evolution of physical systems.
Furthermore, it paved the way for
important developments, such as the discovery of solitons and
 many advancements in the field of  integrable systems and Hamiltonian chaos \cite{zabusky1965interaction,zaslavskiui1972stochastic,campbell2005introduction}. 
More in general,  FPUT chains are a standard tool in many fields, including chemistry and biology~\cite{DNA1,DNA2,berezin2002fermi}.
 
Nowadays,
modern numerical computations have highlighted the fact that the same
initial conditions provided in \cite{fermi1955alamos} can lead to a
thermalized spectrum, characterized by an equipartition of energy
among the independent degrees of freedom, i.e. the Fourier modes~\cite{Ponno2011}. This happens on far larger timescales than those accessible to the original numerical experiment.
Yet,  the problem is important from a fundamental point of view because related to the limit of validity of the statistical mechanics approach. Moreover, it has relevant links to other current research domains as quantum chaos, solid-state physics and Anderson localisation~\cite{berman2005fermi,dauxois2005anti,fishman2012nonlinear,nazarenko2019wave}.
Moreover, in the last decades, important nano-technological applications to heat conduction in low-dimensional materials~\cite{chang2008breakdown,chen2010remarkable,shen2010polyethylene,yang2010violation,liu2012anomalous,chen2021non,huberman2019observation,barbalinardo2021ultrahigh} have motivated a renewed interest in anharmonic chains~\cite{lepri1998anomalous,prosen2000momentum,lepri2003thermal,basile2006momentum,dhar2008heat,wang2020thermal,benenti2020anomalous}. Among other models, the FPUT system has become one of the prototypical benchmarks for anomalous thermal transport in 1D materials~\cite{lepri2003thermal,lepri2016thermal}, as opposed to the expected Fourier's law of heat conduction in 3D solids.


Although many relevant issues have been addressed only recently and many questions remain open, in the last 15 years the field of nonlinear chains has experienced numerous
breakthroughs. In particular, these are due to many new achievements in the physical and
mathematical understanding of the problem of thermalization; notably,
the increasing power of CPUs  has permitted accurate
simulations  at a very weak nonlinearity
\cite{benettin2009fermi,benettin2007fermi,benettin2011time,pistone2018universal}. These results seem to contrast, at least
in the thermodynamic limit, the old idea that there exists a threshold
in energy below which the system does not thermalize
\cite{izrailev1966statistical,livi1985equipartition,casetti1997fermi,deluca1995energy}.
In \cite{benettin2013fermi}, numerical evidence was given that for the 
$\alpha$-FPUT
at small nonlinearity two well separated time scales play a role in the
problem. At small times, the energy is practically shared only by a
small subset of modes; at large times,  energy
equipartition among all modes is reached. The phenomenon observed in
the first time scale is essentially integrability; moreover, it was
noted that for the $\alpha$-FPUT model the reference integrable model
is not the linear chain, but the Toda model \cite{benettin2020understanding,grava2020adiabatic}. 
Furthermore, some studies on the FPUT problem have also pointed out
the role of breathers \cite{flach2005q,gershgorin2005renormalized}, which are non thermal object,
and may play a role in slowing down the thermalization time scale
\cite{Danieli2017intermittent,christodoulidi2010energy,gershgorin2005renormalized}.
These studies are mostly devoted to the understanding of the reasons why Fermi and collaborators did not observe thermalization in the short time they simulated the system. The reason why the 
thermalization is observed in numerical computation (both for small 
number of masses and in the thermodynamic limit) is still debated. 
Recently, the problem has been considered through a different
angle of attack, within the
framework of the so-called {\it Wave Turbulence (WT) theory}
\cite{zakharov2012kolmogorov,nazarenko2011wave,newell2011wave,galtier2022physics}.  
The Wave Turbulence approach has allowed for 
the understanding of the mechanisms underlying the different regimes,
and the estimation of the time scale of
equipartition \cite{PNAS,lvov2018double, pistone2018thermalization,pistone2018universal,
  fu2019universalT,fu2019universal,bustamante2019exact}.
In the nonhomogeneous settings, it has allowed for theoretical predictions of the anomalous exponents of the thermal conductivity observed in the numerical experiments~\cite{pereverzev2003fermi,aoki2006energy,mellet2015anomalous,lepri2016thermal}, and for an interpretation of the anomalous behavior in terms of absence of local thermalization~\cite{dematteis2020coexistence,de2022anomalous}.
Extensions of the Wave Turbulence approach have pointed out also some surprises, such as the existence of
anomalous correlators in a classical chain like the $\beta$-FPUT
system \cite{zaleski2020anomalous}.

Wave Turbulence can be considered as a statistical mechanics of a
large number of interacting waves in the limit of small nonlinearity.
The key object of the theory is the {\it wave kinetic
  equation} that describes the evolution in time of the wave action
spectral density function, which provides the distribution of energy among waves. 
The equation is the analogous of the
Boltzmann equation for a dilute gas, and is also known as the phonon
Boltzmann equation.  The wave kinetic equation allows both for {\it
  out} and {\it in} equilibrium stationary solutions; the latter
corresponds  to the classical thermalization previously
discussed. The out of equilibrium solutions, not discussed in this 
review, are relevant for open systems, where a forcing and a dissipation 
act on different scales; such solutions correspond to the Kolmogorv
spectrum in fluid turbulence, and are characterized by cascades of 
energy or wave action \cite{nazarenko2011wave}. 
The fundamental mechanisms that bring the system into a
stationary state are the resonant interactions among normal 
modes that diagonalize the quadratic part
of the Hamiltonian. 

 Wave Turbulence is a very general framework, which
has been applied with success to a variety of fields, among which
 gravity waves, Bose-Einstein condensates, optical waves, plasma
waves, elastic waves, internal waves in stratified 
fluids, capillary waves, magnetohydrodynamics, 
gravitational waves, Kelvin waves, inertial waves in rotating 
fluids, solid states physics (chains) \cite{zakharov2012kolmogorov,nazarenko2011wave,galtier2022physics, picozzi2014optical}. Many geophysical predictions
are actually based on wave kinetic models.  Not only a large community
of physicists and geo-physicists is active in this research, but the
subject has become an important topic in mathematical-physics and
mathematics. Indeed, despite physicists use the wave kinetic equation 
for applications, its rigorous derivation is still a matter of active 
research: only very few rigorous results have been obtained so far
\cite{aoki2006energy,spohn2006phonon,lukkarinen2016kinetic,buckmaster2020kinetic,staffilani2021wave,deng2021derivation,deng2021full}.
 
The present review is aimed at giving a comprehensive overview of the results obtained for one-dimensional chains like FPUT through the Wave Turbulence theory, thus complementing the excellent books and reviews appeared over the last 30 years~\cite{ford1992fermi,weissert1999genesis,Israilev50,carati2005fermi,gallavotti2008fermi}, which focused on more classical approaches.
We hope to provide also hints on the most interesting questions to be studied in the future.

\label{sec:2} 
\section{A slow start: the harmonic chain}
Before entering the world of nonlinear interactions, we find it useful to start with a 
discussion on linear particle chains and equilibrium statistical mechanics. The following sections will  
also help in introducing the notation that will be used in the 
description of the Wave Turbulence approach. 

\subsection{Hamiltonian formulation and equations of motion}
The Hamiltonian for a chain of $N$ identical particles of mass $m$, subject to a nearest-neighbor harmonic restoring force, can be expressed as the unperturbed Hamiltonian 
\begin{equation}
{\mathcal H}_0=\sum_{j=0}^{N-1}\left(\frac{1}{2 m}p_j^2+\chi\frac{1}{2}(q_j-q_{j+1})^2\right)\,.
\end{equation}
$q_j=q_j(t)$ is the displacement of the particle $j$ from its equilibrium position and $p_j=p_j(t)$ is the associated momentum; $\chi$ is the stiffness constant of the springs. At equilibrium, the inter-particle distance is $a$. The canonical equations of motion are:
\begin{equation}
\begin{split}
&\dot q_j=\frac{\partial {\mathcal H}_0}{\partial p_j}=\frac{p_j}{m}\,,\\
&\dot p_j=-\frac{\partial {\mathcal H}_0}{\partial q_j}=\chi (q_{j+1}+q_{j-1}-2 q_j)\;,
\end{split}
\end{equation}
\begin{equation}
\end{equation}
with $j =0,1,...,N-1 $. Newton's equations reduce to:

\begin{equation}
m\ddot q_j=\chi (q_{j+1}+q_{j-1}-2 q_j)\,.
\label{odes_fput}
\end{equation}
We assume periodic boundary conditions, i.e. $q_N$ = $q_0$, so that we can define the Direct and Inverse Fourier Transforms as follows:
\begin{eqnarray}
Q_k=\frac{1}{N}\sum_{j=0}^{N-1} q_j e^{-i 2\pi k
  j/N},\quad q_j=\sum_{k=-N/2+1}^{N/2} Q_k e^{ i 2\pi j k/N},\nonumber\\
P_k=\frac{1}{N}\sum_{j=0}^{N-1} p_j e^{-i 2\pi k  j/N},\quad p_j=\sum_{k=-N/2+1}^{N/2} P_k e^{ i 2\pi j k/N},\nonumber\\
\label{DFT}
\end{eqnarray}
where $k$ are discrete wavenumbers and $Q_k$ with $P_k$ are the
Fourier amplitudes.

The Fourier transform is a canonical transformation and the equations of motion assume the form:
\begin{equation}
    \dot P_k = -\frac{\partial {{\mathcal H}_0}}{\partial Q_k}\,,
    \ \ 
    \dot Q_k = \frac{\partial {{\mathcal H}_0}}{\partial P_k}\,,
    \label{HamiltonianEquationsOfMotionFourier}
    \end{equation}
where the Hamiltonian takes the following form:
\begin{equation}
{\mathcal H}_0=\frac{N}{2m}\sum_{k=0}^{N-1}\left(|P_k|^2+m^2 \omega_k^2|Q_k|^2\right),
\end{equation}
with 
\begin{equation}
\omega_k^2=4(\chi/m)\sin(\pi k/N)^2.\;
\label{eq:disp_rel^2}
\end{equation}
Newton's equations in Fourier space take the following form:
\begin{equation}
\ddot Q_k+\omega_k^2Q_k=0,
\end{equation}
i.e., a set of decoupled harmonic oscillators.
We introduce an additional transformation
from the variables $P_k(t),\ Q_k(t)$ to the complex variables $a_k=a_k(t)$, as
\begin{equation}\label{NormalMode}
\begin{split}
&a_k=\frac{1}{\sqrt{2 f_k}}(P_k-i f_k Q_k),\\
&a^*_{N-k}=\frac{1}{\sqrt{2 f_k}}(P_k+i f_k Q_k).
\end{split}
\end{equation}
This transformation is canonical if $f_k=f_{N-k}=f^*_{k}$. If we make the choice $f_k=m\omega_k$, where we remark  that $\omega_k$ has been selected as the positive root of Eq.~\eqref{eq:disp_rel^2}, then $a_k$ becomes
the amplitude of the normal mode of the system, known as the {\it normal} variable. The
canonical equations of motion in terms of $a_k$ take
the form:
\begin{equation}i \dot a_k = \frac{\partial {\mathcal H}_0}{\partial a_k^*}~,~\text{where}~
{\mathcal H}_0=N\sum_{k=0}^{N-1}\omega_k |a_k|^2,
  \label{HamiltonianEquationsOfMotionak}
\end{equation}
and the equations of motion become $ i \dot a_k=\omega_k a_k$, whose  analytical solution is
$a_k(t)=a_k(t=0)e^{-i\omega_k t}$.
Finally, the wave action-angle variables can be introduced as follows:
\begin{equation}
\label{action-angle}
a_k=\sqrt{J_k}e^{-i\theta_k }\,,
\end{equation}
where the actions $J_k=J_k(t)$ and the angles $\theta_k=\theta_k(t)$ are real valued and obey the equations of motion:
\begin{equation}
\begin{split}
&\dot\theta_k=\omega_k,\\
&\dot J_k=0,
\end{split}
\end{equation}
which imply that the wave action variables are constants of motion and $\theta_k(t)=\theta_k(t=0)+\omega_k t$. The system is Hamiltonian, with
\begin{equation}\label{eq:H0}
{\mathcal H}_0=N\sum_{k=0}^{N-1}\omega_k J_k\,,
\end{equation}
and the total wave action is given by:
\begin{equation}\label{eq:N}
{\mathcal N} =N\sum_{k=0}^{N-1} J_k\,.
\end{equation}
The total wave action $\mathcal N$ can be interpreted as a total ``number'' of
waves in the system.
\subsection{From discrete to continuum: taking the limit $N\rightarrow \infty$} \label{subsec:longwave}
One of the typical requirements in statistical physics is that the number of particles or masses should be large enough, $N\rightarrow \infty$, to guarantee a proper definition of macroscopic variables. There is no unique way of taking such limit, and, in the following sections, we will consider two ways that have different physical meaning.
\subsubsection{The long-wave approximation} \label{subsec:longwave}
In the framework of particle chains, one way of taking the limit $N\rightarrow \infty$ is to assume that the spacing between the particles at rest goes to zero, $a\rightarrow 0$, in such a way that the chain length $L=Na$ remain constant. Then, 
\begin{equation}
q_j(t)\rightarrow q(x,t),\;\;\;\;\; q_{j\pm1}(t)\rightarrow q(x\pm a,t)
\end{equation}
and the set of ODE's in eq. (\ref{odes_fput}) reduces to a second order in time partial differential equation:
\begin{equation}
\frac{\partial^2 q} {\partial t^2}-v^2 \frac{\partial^2 q} {\partial x^2}=\frac{\chi a^4}{12 m }\frac{\partial^4 q} {\partial x^4}+\mathcal{O}(a^4),
\label{eq:linear_bous}
\end{equation}
where $v^2=\chi a^2/m$, and
 sixth and higher derivatives in space have been neglected.
Note that, while taking the $N\to\infty$ limit, typically one wants to preserve the mass density $\rho=m/a$, and also the propagation speed $v=\sqrt{\chi/m} a$. These further constraints would impose the scalings $m\propto N^{-1}$ and $\chi\propto N$, but they are not the only possible choices.
Equation  (\ref{eq:linear_bous}) can be thought of as a linear Boussinesq equation, i.e. a wave equation corrected with a fourth-order dispersive term. This limit is known as the long-wave approximation; indeed, the same result can be obtained by Taylor expanding directly the dispersion relation 
in (\ref{eq:disp_rel^2}) for small $a$:
\begin{equation}
\omega_k^2= \left(\frac{2 \pi k }{L}\right)^2v^2 - \left(\frac{2 \pi k }{L}\right)^4 \frac{v^2a^2 }{12}+\mathcal{O}(a^4),
\end{equation}
which is nothing but the dispersion relation for the equation (\ref{eq:linear_bous}). The model obtained surely reproduces well the full dispersion relation for small $k$; however, as $k$ becomes large, the accuracy of the model deteriorates. As we will see in section~\ref{subsec:kdv}, in the nonlinear regime, this approach will lead to the Korteweg-De Vries equation which has played an important role in understanding the FPUT numerical observations. 
\subsubsection{The large-box limit or the thermodynamic limit}
The standard approach in statistical mechanics, when studying a gas, is to start with a finite volume and then to take the limit of infinite volume, but keeping the density of the gas constant. In terms of particle chains, the procedure consists of taking $N\rightarrow\infty$ and $L\rightarrow\infty$, keeping their ratio constant, $L/N=a$. This approximation retains the discrete nature of the chain, and, at the same time, the Fourier space becomes dense with $\Delta k=2\pi/L\rightarrow 0$: the discrete wave-number space becomes continuous. 
In this limit the dispersion relation becomes:
\begin{equation}\label{eq:dispFPUT}
\omega_\kappa^2=4\frac{\chi}{m} \sin\left(\frac{a \kappa}{2}\right)^2,
\end{equation}
where $\kappa=2\pi k/L$, and $\kappa\in[0,2\pi/a]$. This limit preserves the shape of the dispersion relation for all wave numbers. 
Anticipating the formalism that will be developed in Section \ref{sec:WT},
we can introduce the main statistical object to be studied, namely the wave-action spectrum, which is defined as
\begin{equation} \label{eq:wave_ac_spec}
 n_\kappa(t)\equiv n(\kappa,t)=\lim_{\Delta k\rightarrow 0}\frac{\langle J_k(t)\rangle}{\Delta k}
 \end{equation}
where  $\langle...\rangle$ is an average procedure over the equilibrium 
distribution.
The equations \eqref{eq:H0} and \eqref{eq:N} take then the following form:
\begin{equation}
{\mathcal H}_0=\frac{L}{a}\int_{0}^{2\pi/a}\omega_{\kappa} n_\kappa d\kappa\,,
\end{equation}
{and the total wave action is given by:}
\begin{equation}
{\mathcal N} =\frac{L}{a}\int_{0}^{2\pi/a} n_{\kappa} d\kappa\,.
\end{equation}
These considerations will be further discussed  when the Wave Turbulence theory will be introduced in Section~\ref{sec:WT}.

\subsection{Some results from classical equilibrium statistical mechanics}
Before entering into the discussion of the out of equilibrium statistical mechanics, we review some of the results that can be achieved by assuming that the system is in equilibrium. 
The discussion that follows has been inspired by recent literature 
\citep{kaufman1986wave,makris2020statistical,rumpf2008transition,rumpf2009stable,
rumpf2007growth,baldovin2021statistical}.


We assume that the system is in thermodynamic equilibrium with a
reservoir which can exchange energy and wave action with the
system. The
thermodynamic quantities that are constrained in the ensemble are the
chemical potential, $\mu$, and the absolute temperature $T$.
In such a setup, it is the reservoir that drives the system to thermal
equilibrium. By contrast, the {\it nonlinear} systems may relax
to thermal equilibrium by themselves, without the need for an external
reservoir. In this section, we keep our focus on {\it linear} systems.
 The joint $N$-mode probability density function (PDF) for the ensemble is given
by
\begin{equation}
P(\{J_k,\theta_k\})= e^{-\beta ({\mathcal H_0} -\mu {\mathcal N})},
\end{equation}
%
with ${\mathcal N}=N\sum_k J_k$ and $1/\beta=k_B T$, where $k_B$ is the Boltzmann constant. 
The mean spectral energy for each mode $k$ in Fourier space is defined as:
\begin{equation}
\langle e_k \rangle^{(eq)}=\frac
{\int_0^{2\pi}... \int_0^\infty \omega_k J_k e^{-N \beta\sum_j (\omega_j-\mu )J_j }dJ_1...dJ_k... dJ_N d\theta_1...d\theta_k...\theta_N} 
{\int_0^{2\pi}... \int_0^\infty e^{-N \beta\sum_j (\omega_j-\mu )J_j}dJ_1...dJ_k... dJ_N d\theta_1...d\theta_k...\theta_N}
\end{equation}
where $e_k = \omega_k J_k$. The integration over angles can be directly
performed, and it is straightforward to show that, using the properties of
the exponentials,  the only integral to be calculated is the
following:
\begin{equation}
\langle e_k \rangle^{(eq)}=\frac
{\int_0^\infty \omega_k J_k e^{-N \beta (\omega_k- \mu)  J_k}dJ_k } 
{\int_0^\infty e^{-N\beta (\omega_k- \mu)J_k}dJ_k}=
\frac{1}{N\beta}\frac{\omega_k}{\omega_k-\mu}
\label{eq:RJenergy_discrete}
\end{equation}
The corresponding distribution of mean  wave action takes the following form:
\begin{equation}
\langle J_k \rangle^{(eq)}=\frac{\langle e_k \rangle}{\omega_k}=
\frac{1}{N}\frac{k_B T}{\omega_k-\mu}.
\label{eq:RJ_discrete}
\end{equation}
This is known as the Rayleigh-Jeans distribution~\cite{zakharov2012kolmogorov}. The same result, in
 the large-box limit, will be obtained as an exact stationary solution
 of the wave kinetic equation (cf. section~\ref{sec:coll-inv}). We note again that the
 {\it nonlinear} chain may relax to equilibrium by itself, whereas the
 {\it linear} chain does need the reservoir. If $\mu=0$, we obtain
 the classical equipartition of energy among the degrees of freedom of
 the system:
\begin{equation}
\langle e_k \rangle^{(eq)}=\frac{1}{N \beta}=\frac{k_B T}{N}.
\end{equation}
This is the result that E. Fermi and collaborators where expecting in their simulations~\cite{fermi1955studies}. 
\subsection{The entropy at equilibrium}
Following the result in \cite{onorato2022equilibrium}, we calculate the Boltzmann entropy for a noninteracting system in the grand canonical ensemble framework, starting from its definition 
\begin{equation}
    S_B(\beta,\mu) = \ln \Omega(\beta,\mu),
    \label{entropy_bol}
\end{equation}
where  $\Omega$ is the number of possible microscopic states. The calculation (see \cite{onorato2022equilibrium} for details), based on the Laplace transform yields the following result:
\begin{equation}
S_B(\beta,\mu) =\ln \Omega(\beta,\mu)=
\sum_{k=0}^{N-1}
\ln \langle J_k \rangle^{(eq)}.
 \label{eq:entropy_discrete}
\end{equation}
Note that the argument of the logarithmic function is the Rayleigh-Jeans distribution in Eq.~\eqref{eq:RJ_discrete}.  Interestingly, it will turn out that, using the Wave Turbulence approach, 
an entropy can also be defined in out-of-equilibrium conditions; such entropy matches Eq.~\eqref{eq:entropy_discrete} with  the only difference that the argument of the logarithm, $ \langle J_k \rangle$, can be computed also in out of equilibrium conditions.

We now consider the Gibbs entropy at fixed $k$ defined as:
\begin{equation}
S_G=-\int_0^{\infty}p(J_k) \ln p(J_k) dJ_k,
\label{entropy_reduced}
\end{equation}
where $p(J_k)$ is the $1$-mode action PDF. We maximize the Gibbs entropy with respect to $p(J_k)$, subject to the following constraints (see \cite{kaufman1986wave,makris2020statistical}):
\begin{equation}
\langle J_k\rangle =\int_0^{\infty} J_k p(J_k) dJ_k,
\end{equation}
\begin{equation}
1 =\int_0^{\infty}  p(J_k) dJ_k.
\end{equation}
The maximization procedure 
leads to
\begin{equation} \label{eq:sol_pdf}
p(J_k)=\frac{1}{\langle J_k \rangle }\exp \left[{-\frac{J_k}{\langle J_k \rangle }}\right].
\end{equation}
This result, obtained in \cite{kaufman1986wave}, is the well-known exponential distribution for the wave actions. It will be shown that this is the maximal-entropy stationary solution of the evolution equation for the PDF for the wave action, see Section \ref{sec:WT}.

\label{sec:3} 

\section{Nonlinearity in action: Anharmonic chains}
We now turn the attention to the deterministic microscopic evolution and discuss the effect of nonlinearity.
We start from the general Hamiltonian of a one
dimensional chain of $N$ identical atoms interacting nonlinearly:
  \begin{equation}\label{eq:6.1}
{\mathcal H}=\sum_{j=0}^{N-1} \left[\frac{1}{2 m}p_j^2+V(q_j,q_{j+1})\right],
\end{equation}
where $V(q_j,q_{j+1})$ is a potential limited to nearest-neighbour interaction.
The potential may assume different forms. In the present review, we will discuss the following models:
%
%
%
\begin{itemize}
\item $\alpha$-FPUT:
\begin{equation}\label{eq:alphaFPUT}
V_\alpha=\frac{\chi}{2}(q_j-q_{j+1})^2+\frac{\alpha}{3}(q_j-q_{j+1})^3,\;\;\;\;\;\;
\end{equation}
where $\alpha$ is the nonlinearity parameter. The equations of motion are:
\begin{equation}
m\ddot q_j=
\chi \left(q_{j+1}+q_{j-1}-2q_j\right)+
\alpha \left[ (q_{j+1}-q_{j})^2 - (q_i-q_{i-1})^2\right];
\label{eq:alpha}
\end{equation}

\item
$\beta$-FPUT model:
\begin{equation}\label{eq:betaFPUT}
V_\beta=\frac{\chi}{2}(q_j-q_{j+1})^2+\frac{\beta}{4}(q_j-q_{j+1})^4,\;\;\;\;\;\;
\end{equation}
where $\beta$ is the nonlinearity parameter. This leads to 

\begin{equation}
m\ddot q_j=
\chi \left(q_{j+1}+q_{j-1}-2q_j\right)+
\beta \left[ (q_{j+1}-q_{j})^3 - (q_i-q_{i-1})^3\right];
\label{eq:beta}
\end{equation}

\item $\alpha+\beta$-FPUT model: it combines the nonlinearity of the $\alpha$- and $\beta$-FPUT in a single model.

\item Discrete  Nonlinear  Klein Gordon (DNKG) chain:
\begin{equation}
V_{\rm KG}=\frac{\chi}{2}(q_j-q_{j+1})^2+\frac{s}{2}q_j^2 +\frac{1}{4} g q_j^4,\;\;\;\;\;\;
\end{equation}
This potential leads to the DNKG system of equations:
\begin{equation}
m\ddot q_j=
\chi \left(q_{j+1}+q_{j-1}-2q_j\right)-sq_j-
g q_j^3,
\label{eq:KG}
\end{equation}

\item Toda lattice:
\begin{equation}\label{eq:todapot}
V_{T}=\frac{\chi^3}{4 \alpha^2}\left[\exp\left(\frac{2 \alpha}{\chi} (q_j-q_{j+1})\right)- \frac{2\alpha}{\chi}  (q_j-q_{j+1}) -1 \right],
\end{equation}

\begin{equation}
m \ddot{q}_j = \frac{\chi^2}{2\alpha}\left\{\exp\left[\frac{2 \alpha}{\chi} (q_{j+1}-q_j)\right]-\exp\left[\frac{2 \alpha}{\chi}(q_j-q_{j-1})\right]\right\}.
\label{eq:toda}
\end{equation}
\end{itemize}

\subsection{Equations in normal variables} \label{subsec:normal_variable}
For completeness, here below, we report the Hamiltonians and the evolution equations in Fourier space, 
assuming periodic boundary conditions,  in the {\it normal} 
variables defined in Eq. (\ref{NormalMode}). For simplicity, from now on we will set $m=\chi$=1. All the sums here 
below are intended from 0 to $N-1$ and we use the following notation for the Kronecker delta:
\begin{equation}
	\label{delta1,2+3}
	\delta_{a,b,c,...}^{l,m,n...}=
	\begin{cases}
		1 & \text{if $k_a+k_b+k_c+...=k_l+k_m+k_n+... \pmod N$}\,, \\
		0 & \text{otherwise}\,,
	\end{cases}
\end{equation}

\paragraph{$\alpha$-FPUT}
\begin{equation}
\begin{split}
&\frac{\mathcal H}{N}=\sum_k \omega_k|a_k|^2+\alpha\sum_{k_1,k_2,k_3} \bigg[
\frac{1}{3}\left( V_{1,2,3}a_1a_2a_3+c.c.\right)\delta_{1,2,3}+\\
&+ \left(V_{-1,2,3}a_1^*a_2a_3+c.c.\right)\delta_{1}^{2,3}\bigg],
\end{split} \label{eq:Ham_alpha}
\end{equation}
\begin{equation}
\begin{split}
i \frac{d a_1}{dt}=\omega_1 a_1+\alpha \sum_{k_2,k_3} \bigg[
 V_{-1,2,3}a_2a_3\delta_{1}^{2,3}-2 V_{-3,1,2}a_2^*a_3\delta_{1,2}^3-V_{1,2,3}a_2^*a_3^*\delta_{1,2,3}\bigg].
\end{split}
\end{equation}
The analytical form of the matrix $V_{1,2,3}=V_{k_1,k_2,k_3}$ is reported in \cite{bustamante2019exact}.
\paragraph{$\beta$-FPUT}
\begin{equation}
\begin{split}
&\frac{\mathcal H}{N}=\sum_k \omega_k|a_k|^2+\beta \sum_{k_1,k_2,k_3,k_4} \bigg[
\frac{1}{4}\left(T_{1,2,3,4} a_1a_2a_3a_4+c.c.\right)\delta_{1,2,3,4}+\\
&+T_{-1,2,3,4} \left(a_1^*a_2a_3a_4+c.c.\right)\delta_{1}^{2,3,4}
+\frac{3}{2}T_{-1,-2,3,4} \left(a_1^*a_2^*a_3a_4+c.c.\right)\delta_{1,2}^{3,4}\bigg]
\end{split}\label{eq:Ham_beta}
\end{equation}
\begin{equation}
\begin{split}
&i \frac{d a_1}{dt}=\omega_1 a_1+\beta \sum_{k_2,k_3,k_4} \bigg[
 T_{-1,2,3,4}a_2a_3a_4\delta_{1}^{2,3,4}+3 T_{-1,-2,3,4}a_2^*a_3a_4\delta_{1,2}^{3,4}+\\
 &+3T_{-4,1,2,3}a_2^*a_3^*a_4\delta_{1,2,3}^{4}+
 T_{1,2,3,4}a_1^*a_2^*a_3^*a_4^* \delta_{1,2,3,4} \bigg]
\end{split}
\end{equation}
The analytical form of the matrix $T_{1,2,3,4}=T_{k_1,k_2,k_3,k_4}$ is reported in \cite{bustamante2019exact}	
\paragraph{DNKG}
\begin{equation}
\begin{split}
&\frac{\mathcal H}{N}=\sum_k \omega_k|a_k|^2+g \sum_{k_1,k_2,k_3,k_4} \bigg[
\frac{1}{4}\left(W_{1,2,3,4} a_1a_2a_3a_4+c.c.\right)\delta_{1,2,3,4}+\\
&+W_{-1,2,3,4} \left(a_1^*a_2a_3a_4+c.c.\right)\delta_{1}^{2,3,4}
+\frac{3}{2}W_{-1,-2,3,4} \left(a_1^*a_2^*a_3a_4+c.c.\right)\delta_{1,2}^{3,4}\bigg],
\end{split}\label{eq:Ham_DNLKG}
\end{equation}
\begin{equation}
\begin{split}
&i \frac{d a_1}{dt}=\omega_1 a_1+\beta \sum_{k_2,k_3,k_4} \bigg[
 T_{-1,2,3,4}a_2a_3a_4\delta_{1}^{2,3,4}+3 T_{-1,-2,3,4}a_2^*a_3a_4\delta_{1,2}^{3,4}+\\
 &+3T_{-4,1,2,3}a_2^*a_3^*a_4\delta_{1,2,3}^{4}+
 T_{1,2,3,4}a_1^*a_2^*a_3^*a_4^* \delta_{1,2,3,4} \bigg],
\end{split}
\end{equation}
where
 \begin{equation}
 \label{eq:disp_rel_DNLKG}
 \omega_k=\sqrt{1+4 \sin(\pi k/N)^2}.
 \end{equation}
The analytical form of the matrix $W_{1,2,3,4}=W_{k_1,k_2,k_3,k_4}$ is reported in \cite{pistone2018universal}


\subsection{The $\alpha$-FPUT and the Korteweg de Vries Equation}
\label{subsec:kdv}
In the sixties, when the report by FPUT  was finally available to the scientific community, many attempts to understand the surprising recurrent dynamics were made.
One of the most important, most notably for the long-standing impact on mathematical-physics and physics, was given by Zabuski and Kruskal who decided to study the problem in its continuous limit~\cite{zabusky1965interaction}.
The procedure used  is the same shown in Section \ref{subsec:longwave}, where the long wave approximation is taken. The nonlinearity is also now included in the calculation in such a way that dispersion and nonlinearity balance each other at their leading order; for the $\alpha$-FPUT lattice, the resulting equation is the Boussinesq equation:
\begin{equation}
\frac{\partial^2 q} {\partial t^2}-v^2 \frac{\partial^2 q} {\partial x^2}=\frac{\chi a^4}{12 m }\frac{\partial^4 q} {\partial x^4}+ \frac{2a^3\alpha}{m} \frac{\partial^2 q} {\partial x^2}\frac{\partial q} {\partial x}+
\mathcal{O}(a^6, a^5 \alpha)~,
\label{eq:bous}
\end{equation}
where the expansion is truncated at the leading nonlinear and dispersive terms. It is worth noting that the limit $N\rightarrow \infty$ and $a\rightarrow 0$ is generally not unique and one could get other equations balancing nonlinearity and dispersion in a different way.
Assuming that waves propagate only in the positive $x$ direction, and defining $\eta(x,t)=\partial q(x,t)/\partial x$, the equation \eqref{eq:bous} becomes the Korteweg de Vries (KdV) equation~\cite{zabusky1965interaction,dauxois2006physics}:
\begin{equation}
\frac{\partial \eta} {\partial t}+6\alpha \eta \frac{\partial \eta} {\partial x}+
\frac{\partial^3 \eta} {\partial x^3}=0,
\label{eq:kdv}
\end{equation}
where variables have been scaled in order to get the equation in standard form, see for detail \cite{ablowitz2011nonlinear}.
This is historically one of the most fundamental equations for the nonlinear theories in dimension 1D+1.

Zabuski and Kruskal \cite{zabusky1965interaction} performed numerical simulations of the KdV equation showing two relevant phenomena for the interpretation of the numerical results in \cite{fermi1955studies}:  the first one is the observation of the recurrence, just like in the numerical experiments of the FPUT system, and the second one is the the observation of solitons. The authors coined  the word ``solitons'' to indicate the peculiar solution of the equation which behaved as a quasi-particle with localised and finite energy, displaying a rigid translation movement. From a physical point of view, solitons are solutions corresponding to an exact balance between nonlinear and dispersive effects.

These results paved the way for extensive developments in mathematical-physics. Integrable PDEs were discovered and  the Inverse Scattering Transform was developed as a tool for studying solutions of such equations  
\citep{faddeev1959inverse,faddeyev1963inverse,gardner1967method,lax1968integrals,zakharov1971korteweg,shabat1972exact,zakharov1974scheme,ablowitz1973nonlinear,ablowitz1973method}.


Following the work of Zabusky and Kruskal, in \cite{ponno2003soliton}, the FPUT problem was reconsidered from the point of view of soliton theory. For a special class of long-wavelength initial data, it has been shown that the energy spectrum observed in the early stages of the $\alpha$-FPUT computations is determined by solitons~\cite{ponno2003soliton}, and such spectrum does not correspond to equipartition of energy. Some years later Bambusi and Ponno~\cite{bambusi2006metastability} showed that a pair of KdV equations constitute the resonant normal form of the $\alpha$-FPUT system in the long-wave approximation (cf. section~\ref{sec:3.3}). In~\cite{bambusi2006metastability}, such a normal form was used to prove the existence of a metastability phenomenon; in particular, it was shown that the time average of the energy spectrum rapidly settles to a distribution corresponding to a packet of low wavenumber modes, which remains unchanged up to the time scales of validity of the approximation. The effects of higher order dynamics in the KdV hierarchy and the asymptotic integrability properties of $\alpha$-FPUT  is discussed in \cite{gallone2021korteweg}.

As a general comment, we mention that the solitons of the KdV equation can explain the short time dynamics of the  $\alpha$-FPUT for initial data  concentrated at low wave-numbers; however, they cannot explain the full rich dynamics of the anharmonic chains, which has been unveiled by more recent numerical simulations \cite{benettin2011time}. The reason is that the KdV equation is obtained as a long-wave approximation of the $\alpha$-FPUT, and the two linear dispersion relations agree only for low wave-numbers. Any truncated long-wave expansion of $\alpha$-FPUT would fail to properly describe the phenomenon of thermalization characterized by equipartition of energy among all modes (including the highest ones). 

With the renewed interest triggered by the work by Zabusky and Kruskal~\cite{zabusky1965interaction},  soliton dynamics and the FPUT recurrence have been observed in many fields of physics, including plasma, solid state, biological systems \cite{infeld2000nonlinear,ashcroft2022solid,davydov1973theory,putnam1981resonant,peyrard1989statistical,dauxois2006physics,gariaev1991holographic,guasoni2017incoherent,Trillo2016}.


\subsection{The long-wave expansion of the $\beta$-FPUT: the modified KdV equation}\label{sec:3.3}
The $\beta$-FPUT chain is characterized by a cubic nonlinearity in the equations of motion; therefore, its dynamics  cannot be described by the standard KdV equation. However, the linear dispersion relation of the $\beta$-FPUT is identical to the $\alpha$-FPUT  one. Therefore, as for for the  $\alpha$-FPUT, it is possible to perform a long-wave expansion. The procedure has been exploited in \cite{driscoll1976explanation}, where the limits $N\rightarrow\infty$ and $a\rightarrow 0$ were taken, and, assuming waves traveling in only one direction,  the modified KdV equation (mKdV) is obtained:
\begin{equation}
\frac{\partial \eta} {\partial t}+6 \beta \eta^2 \frac{\partial \eta} {\partial x}+
\frac{\partial^3 \eta} {\partial x^3}~,
\label{eq:mkdv}
\end{equation}
where, again, $\eta(x,t)$ is related to the spatial derivative of the displacement of the particles (in the continuum limit) with respect to the equilibrium position and the variables have been suitably scaled.   The mKdV equation is integrable through Inverse Scattering Transform~\cite{wadati1973modified}.  It has been shown that solutions of the mKdV equation are unstable ($\beta>0$), and that these mKdV instabilities correspond to the instabilities observed on the lattice~\cite{driscoll1976explanation}. It was then conjectured that those instabilities may be important in determining whether or not a system behaves stochastically.
In the framework of the mKdV equation, the FPUT recurrence was studied in details in \cite{pace2019beta}, both for positive and negative $\beta$. 
As previously mentioned for the $\alpha$-FPUT and KdV, the present approach, despite very rich and relevant for understanding the recurrence and the transition to a more complex dynamics, cannot fully describe the thermalization process to an equipartition of energy as expected by Fermi and collaborators~\cite{fermi1955studies}. 

\subsection{The narrow-band approximation of the $\beta$-FPUT:   the Nonlinear Schr\"odinger equation}
As in many other fields of physics, the Nonlinear-Schr\"odinger (NLS) equation plays also an important role in the understanding the phenomenon of  recurrence and the transition to a stochastic dynamics observed in numerical simulations.  Indeed, the NLS equation can be derived for any dispersive wave system with cubic nonlinearity, under the assumption that most of the energy is concentrated around a single wave number, $k_0$. The idea is straightforward~\cite{berman1984limit}: using the normal variables defined in \eqref{NormalMode} in the equations of motion for the $\beta$-FPUT, \eqref{eq:beta}, and expanding 
$\omega_k$ around $k=k_0$, to second order in the linear part of the equation and to the leading order nonlinear one, after some rescaling and redefinition of the variables, the following  equation can be derived, \cite{berman1984limit}:
\begin{equation}
i \frac{d a_1}{d t}=-\gamma K^2 a_1+\beta T_0 \sum_{K_2,K_3,K_4}
 a^*_2a_3 a_4 \delta_{K_1+K_2, K_3+K_4}
 \label{eq:NLS_F}
\end{equation}
with $a_i=a(K_i,t)$, $K=k-k_0$,  $\gamma=\frac{\pi^2}{N^2} \sin\left[\frac{\pi k_0}{N}\right]$
and $T_0=3 \omega_{k_0}^2/4$. Equation \eqref{eq:NLS_F} is the NLS equation written in 
Fourier space. The NLS equation is integrable through Inverse Scattering Transform, \cite{shabat1972exact,ablowitz1974inverse}; it is well known that the equation has exact soliton 
(bright or dark, depending on the regime, focusing or defocusing) and 
breather solutions \cite{flach2008discrete}. In the focusing case, a plane wave solution can be unstable to side band perturbations,
and the instability is known as modulational instability or Benjamin-Feir instability \cite{yuen1978relationship,sulem2007nonlinear,zakharov2009modulation}.
Being integrable, the NLS has the property of exhibiting a phenomenon of recurrence similar to the one observed by Fermi and collaborators~\cite{berman1984limit,kuznetsov2017fermi,dauxois2005anti}.

\subsection{The Toda  and the FPUT  lattices} \label{subsec:Toda}
In \cite{benettin2013fermi}, the authors performed a numerical study of the 
$(\alpha+\beta)$ FPUT model; see also \cite{ponno2011two} for $\alpha$-FPUT simulations. The paper contains two important considerations: i) for small specific energies  two well-separated time scales are present in the process of thermalization: in the first one, a metastable state is observed, while in  the second one, which is much larger, a statistical equilibrium is reached;  ii) the FPUT numerical results should be interpreted, in the limit of small specific energy and for time up to the metastable state, as a perturbation of the Toda lattice \cite{toda1967vibration}, which is an integrable system through the Inverse Scattering Transform \cite{flaschka1974todaa,flaschka1974todab}.  More specifically, the authors have shown numerically that the dynamics of the $(\alpha+\beta)$ FPUT  for the former time scale is indistinguishable from the Toda dynamics. 

From a mathematical point of view, one can Taylor-expand the Toda potential in eq. (\ref{eq:todapot}) for small $\alpha$ to obtain
\begin{equation}
V_{T}\simeq\frac{\chi}{2} (q_j-q_{j+1})^2+ \frac{\alpha}{3} (q_j-q_{j+1})^3+\frac{\alpha^2}{6 \chi} (q_j-q_{j+1})^4+O(\alpha^3).
\end{equation}
Comparing such potential with the $(\alpha+\beta)$-FPUT,  
\begin{equation}
V_{\alpha+\beta}=\frac{\chi}{2} (q_j-q_{j+1})^2+ \frac{\alpha}{3} (q_j-q_{j+1})^3+\frac{\beta}{4} (q_j-q_{j+1})^4,
\end{equation}
it is evident that the $(\alpha+\beta)$-FPUT and the Toda lattice are tangent, their distance being determined by the difference in the fourth-order pre-factors: $|\beta/4-\alpha^2/(6 \chi)|$.
\begin{figure}\label{fig:Toda_FPUT}
\begin{center}
\includegraphics[width=0.85\columnwidth]{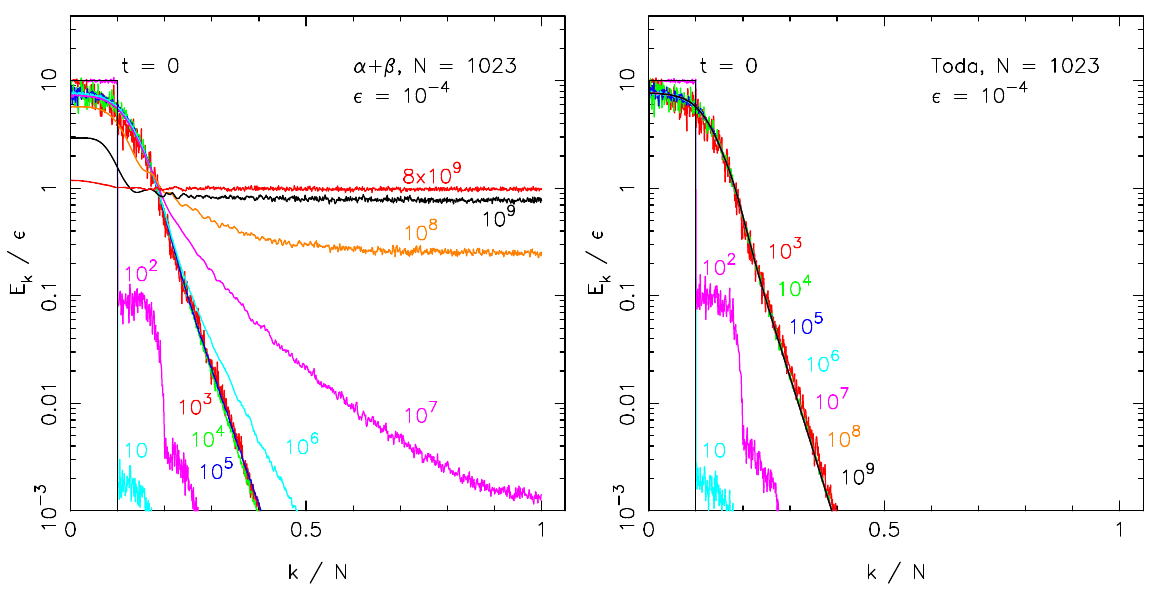}
\caption{Plot taken from reference \cite{Benettin2023FPUmodel}.  Energy spectrum plotted as a function of $k/N$ at selected times. Left: FPUT simulations; Right: Toda simulations. See text for details. }
\end{center}
\end{figure}
In Figure \ref{fig:Toda_FPUT}, taken from \cite{Benettin2023FPUmodel}, we report the results of a numerical computation obtained with the $(\alpha+\beta)$-FPUT and the Toda lattice. Simulations are performed with $N$ = 1023 particles, with $\alpha=1$, $\beta=2$, $\chi=1$, using fixed boundary conditions and the time-averaged harmonic energy density defined as 
\begin{equation}
\bar E_k=\frac{1}{T}\int_0^T\frac{1}{2}\left(P_k^2+\omega_k^2 Q_k^2 \right)dt.
\end{equation}
 At time $t=0$, the  energy density  was equidistributed among the first 10\% of normal modes with random phases (besides time averaging, an ensemble average procedure with 24 realizations with different set of random phases has been applied).  The total initial energy, indicated with $\epsilon$ in the plot,  was set to $10^{-4}$, while the initial profile is the black rectangle. The two plots show that up to time approximately of $t\sim10^5$ the dynamics of the FPUT and Toda lattice are  indistinguishable. After that, the energy spectrum for the Toda lattice remains frozen, while for the FPUT the equipartition of energy among Fourier modes is gradually achieved. 

The idea that the Toda lattice is tangent to FPUT has been exploited theoretically in 
\cite{grava2020adiabatic}, where the authors proved that  with periodic boundary conditions and for large $N$, the first $m$, with $m\ll N$,  integrals of motion of the Toda chain are adiabatic invariants of FPUT, i.e. they are approximately constant along the dynamics of the FPUT. This result followed  some numerical computations on the same problem previously performed and reported in \cite{goldfriend2019equilibration}.



We emphasise that the FPUT problem has fostered the development of the theory of nonlinear integrable Hamiltonian system, a very rich and relevant field.
This approach has been useful to understand some features of the original FPUT problem, most notably in relation to the initial recurrence. 
Yet, it does not tell the whole story, in particular in the long time scales when the system is far from integrability.
In the following section, we review another parallel approach to deal with quasi-integrability and chaos, that is the perturbation theory, including the advances due to Kolmogorov, Arnold, Moser and Nekhoroshev. Beside its fundamental importance for the whole classical mechanics, these developments have also given some insights on the FPUT problem, notably for small system size.
Then, we shall turn to the core of this review, Wave Turbulence theory, in Section~\ref{sec:WT}.


\section{FPUT and its relation to the KAM theorem and to Birkhoff normal forms}
\label{sec:4} 


In order for the reader to better appreciate some of the mathematical results obtained on the FPUT problem, we find it useful to summarize, in a physics-based approach, some key instruments developed for analyzing quasi-integrable systems: the KAM, the Nekhoroshev theorems and the Birkhoff normal forms, see  \cite{arnold2006mathematical,giorgilli2022notes}.

One year before the FPUT report \cite{fermi1955studies} was published,  Kolmogorov, Arnold and Moser formulated what we now know as the KAM theorem \cite{kolmogorov1954conservation,arnold1963proof,moser1962invariant}. The result,  developed for Hamiltonian dynamical systems that are nearly integrable, is related to the persistence of quasi-periodic motions under small perturbations. The whole idea is based on the development of a perturbation theory that partially resolves the problem of small divisors {{(see Section~\ref{sec:KAM} below)} while ensuring the convergence of the expansion. One of the important consequences of the KAM theorem is that, for a large set of initial conditions, the motion remains quasi-periodic. Clearly, such result has attracted the attention of the scientists working on the recurrence phenomenon originally observed numerically in \cite{fermi1955studies}; indeed, the FPUT problem may  be seen as a perturbation of an integrable system. However, such interpretation is not so straightforward: it took  about 50 years since the FPUT original numerical computations for some rigorous results on the applicability of the KAM theorem to the FPUT problem to be achieved. 

\subsection{KAM theorem}\label{sec:KAM}
For a rigorous statement in its most basic form, and for a detailed proof, we refer the reader for example to \cite{poschel2009lecture} or \cite{benettin1984proof}.
The KAM theorem considers  an unperturbed $N$-dimensional integrable Hamiltonian system with ${\mathcal H}_0(\{J_k\})$,  where  $\{J_k\}=J_1,J_2,..,J_k,..,J_N$ denotes the $N$ action variables. 
Being integrable, the phase space is characterized by invariant tori, $J_k=const$, and the motion on the tori is quasi-periodic described by $N$ angle variables, $\theta_k(t)$, which evolve as $\theta_k(t)=\theta_k(t=0)+\omega_k t$, with frequencies $\omega_k=\omega(J_k)=\partial {\mathcal H}_0/\partial J_k$. 
These tori can be {\it resonant} if the frequencies are rationally dependent, i.e., if there exist integers $m_1, m_2, ... , m_N$, not all of which are zero,  such that
\begin{equation}\label{eq:resonances}
m_1\omega_1+m_2\omega_2+...+m_N\omega_N=  0.
\end{equation}
The tori for which such  integers do not exist (frequencies are rationally independent, $\sum m_i \omega_i\ne0$), are known as {\it nonresonant}. The left-hand side of equation (\ref{eq:resonances}) enters explicitly in the denominator when developing the perturbation theory \cite{arnold2006mathematical}; therefore, it plays an important role in establishing the applicability of the KAM theorem.  One requirement for the  applicability of the KAM theorem is that the system is {\it nondegenerate},   i.e., the frequencies are functionally independent:
\begin{equation} \label{eq:nondegenerate}
\det\left(\frac{\partial \omega_k}{\partial J_i}\right)
=\det\left( \frac{\partial^2 {\mathcal H}_0} {\partial J_i \partial J_k}\right)\ne0.
\end{equation}
In a nondegenerate system the {\it nonresonant} tori form a dense set of full measure, while the {\it resonant}  ones form still a dense set yet of measure zero. 

The KAM theorem deals with a perturbation of the integrable Hamiltonian ${\mathcal H}_0$:
\begin{equation}\label{eq:pert_kam}
\mathcal H(\{J_k,\theta_k\})=\mathcal H_0(\{J_k\})+\epsilon \mathcal H_1(\{J_k,\theta_k\}),
\end{equation}
with $\epsilon\ll1$.
 The theorem describes the effect of the perturbation on the {\it nonresonant} tori. Loosely speaking, it states that, if the unperturbed system is nondegenerate and
 for sufficiently small $\epsilon$, then most of the {\it nonresonant} tori survive, and are only slightly deformed. The dynamics on these tori is quasi-periodic with a frequency equal to the unperturbed one. The smaller the perturbation, the larger the portion of the {\it nonresonant } tori which are preserved.

The classical proof of the theorem is based on perturbation theory. The idea is to make a symplectic coordinate transformation, $\{J_k,\theta_k\} \to \{I_k,\phi_k\}$, such that in the new coordinates the term of order $\epsilon$ in the perturbed Hamiltonian is independent of the phases. The new Hamiltonian will contain the phases at order $\epsilon^2$. The procedure is then iterated to ``push'' the phases to higher order  in $\epsilon$. From a physical point of view, this corresponds to separating the fast evolution of the phases from the slow dynamics of the actions.   The convergence of the perturbation series, assured by the strong nonresonant condition of the frequencies, is the core of the beautiful technical results obtained in the KAM theorem and is known as super-convergence~\cite{arnold2006mathematical, poschel1982integrability,chierchia1982smooth,gallavotti2001meccanica}.
 \subsection{Nekhoroshev theorem}
A fundamental result, somewhat complementary with respect to the KAM
theory, has been obtained by Nekhoroshev
\cite{nekhoroshev1971behavior,nekhoroshev1977exponential}.  
Like the KAM theorem, Nekhoroshev's theorem applies to quasi-integrable
Hamiltonians of the form (\ref{eq:pert_kam}), if the unperturbed Hamiltonian  $\mathcal H_0$ presents some non-degeneracy property, see~\cite{nekhoroshev1977exponential,nekhoroshev1979exponential}.  The theorem essentially states that, if $\epsilon$ is sufficiently small,
any motion $\{J_k(t), \theta_k(t)\}$ of the system satisfies
\begin{equation}
\vert J_k(t)- J_k(0) \vert \le C \epsilon^a~~{\rm for}~~
0\le t \le T \exp\left(\frac{1}{\epsilon^b}\right),
\end{equation}
where $ a,b,C,T$ are positive constants that depend on the generic properties of the unperturbed Hamiltonian. 
In particular, $a$ and $b$ depend on the number of degrees-of-freedom~\cite{nekhoroshev1979exponential,lochak1992estimates,poschel1993nekhoroshev,guzzo2016steep}. The theorem holds in the regions in which the frequencies of the unperturbed motion satisfy a no-resonance relation at order up to $1/\epsilon$.
The applicability of Nekhoroshev's theorem implies the existence of invariant tori \cite{morbidelli1995connection};  their diffusion speed
is bounded by a superexponential of the inverse of the distance from an invariant torus. 
 \subsection{Normal forms}
The concept of Birkhoff normal form, developed in the framework of the small-oscillation theory near equilibrium points~\cite{birkhoff1927dynamical},
is strictly related to the perturbation theory just described: the objective is to ``push'' the angles as far as possible to higher order in the perturbation parameter $\epsilon$, so that the Hamiltonian, up to some order, depends only on the actions.  An Hamiltonian $H$ is in Birkhoff normal form of degree $2r$, if it can be written as follows:
\begin{equation}
H(\{J_k,\theta_k\})=P_1+P_2+....+P_{r}+H_{r+1/2},
\end{equation}
where $P_j$ are  homogeneous polynomials of degree $j$ in actions which do not depend on phases;
 $H_{r+1/2}$  stands for arbitrary terms of order higher than $r$ in actions, and include also phases.
If $H_{r+1/2}$ is discarded then the Hamiltonian is completely canonically integrable. It is clear that the possibility of finding coordinates such that the Hamiltonian takes the form of a sum of polynomials only function of actions hinges upon the non existence of resonances. However, an asymptotically integrable Hamiltonian may also be written in presence of resonances. Let's assume that at some order resonances exist; then a {\it resonant} normal form is obtained if the Hamiltonian, written in terms of angle-action variables, depends only on the angles associated with resonances. Such form can be completely integrable. 
For a rigorous introduction to Birkhoff normal forms the reader is invited to look at~\cite{bambusi2014introduction}.

\subsection{Applicability to FPUT}
Here below, we report on the main mathematical results concerning the  applicability of KAM and Nekhoroshev theorems to FPUT chains.
A nice gentle discussion on these aspects can be found in scholarpedia \cite{Rink:2009}.

As can be seen from the Hamiltonians in (\ref{eq:Ham_alpha}), (\ref{eq:Ham_beta}) and (\ref{eq:Ham_DNLKG}), a direct application of the KAM theorem is not feasible because the unperturbed Hamiltonians do not satisfy the nondegenarcy condition, eq. (\ref{eq:nondegenerate}), as they depend linearly on actions, $\mathcal H_0=\sum \omega_k J_k$; therefore, more sophisticated work has to be done. 
An important first step in the study of the applicability of  KAM theorem to FPUT was made by Nishida~\cite{nishida1971note},
who conjectured the existence of a completely integrable finite-order Birkhoff normal form, approximation of the original Hamiltonian, in the case of a symmetric potential (the $\beta$-FPUT).
Remarkably, the  proof of this conjecture has been obtained only 35 years later, reported in the fundamental work by Rink \cite{rink2006proof}.
As stated in~\cite{rink2006proof}, it is proven that, thanks to the existence of discrete symmetries~\cite{rink2001symmetry,rink2002direction},  an integrable normal form exists. It is also shown that for the $\beta$-lattice the approximation is nondegenerate in the KAM sense, proving the existence of a large measure of quasi-periodic motions.
This result was improved in ~\cite{henrici2008results}, where the authors prove that any periodic FPUT chain (including odd energy potential) has a normal form approximating the original Hamiltonian.
They prove that, for an odd number of masses $N$, the normal form is of the Birkhoff kind, that is depending only on actions, and nondegenerate. In this case, the KAM results apply straightforwardly. Quite interestingly, for  $N$ even, the normal form is resonant but completely integrable, such that it is possible ultimately to find a non-trivial Birkhoff normal form in all cases. As a comment to the works just mentioned, we report a sentence included in the PhD thesis of A. Henrici \cite{henrici2008normal}:
``Even though we thus rigorously confirm the long-standing conjecture that
the KAM theorem can be applied to FPU chains, it seems unlikely that this
already ``explains'' the FPU paradox, since it remains unclear whether the energy
levels and initial conditions chosen by Fermi, Pasta, and Ulam fit into the ``scheme'' of the KAM theorem. In particular, since the admissible energy levels
for our application of KAM to FPU appear to be becoming smaller and smaller
as the number of particles tends to infinity, it seems rather unlikely that the
KAM theorem is sufficient for the desired explanation of the FPU paradox.''.
The work by Henrici and Kappeler~\cite{henrici2008results} has furthermore a deep significance on our present understanding of the FPUT dynamics. As also investigated later~\cite{henrici2009nekhoroshev,kappeler2009resonant},  their work nicely shows that generic FPUT chains admit an integrable behaviour since they already contain an integrable system, which is the Toda lattice.
While FPUT chains have been  considered as a perturbation of the linear dynamics, these mathematical results indicate that it is often more relevant to consider the FPUT as a perturbation of the Toda lattice.
This point of view was actually pioneered in an important work focused on the computational approach~\cite{ferguson1982nonlinear}, and, as discussed in Section \ref{subsec:Toda},   it has been since corroborated by exhaustive numerical simulations \cite{benettin2013fermi}, and used to find interesting mathematical developments~\cite{bambusi2015some,bambusi2016birkhoff}.

As mentioned, one of the major questions that remains to be answered is the relevance of the mentioned results in the limit {$N\rightarrow \infty$}, since no explicit bounds on the stability of the integrable approximation with $N$ is provided. 
As far as we are aware of, the rigorous applicability of KAM theorem to the Discrete Nonlinear Klein-Gordon system has not been addressed yet.

\subsection{Resonances in the FPUT problem}
A key concept in the KAM theorem is related to the existence of resonances. In the case of the FPUT problem, the condition on frequencies in eq. (\ref{eq:resonances}) is, by construction, combined  with an analogous condition on wavenumbers (the Kronecker Deltas in equation (\ref{eq:Ham_alpha}), (\ref{eq:Ham_beta}) and (\ref{eq:Ham_DNLKG})), so that the general resonances between $M$ waves must satisfy the following two equations:
\begin{equation}
\begin{split}
&k_1+k_2+...+k_S=k_{S+1}+...+k_{M}\;\;\;\; ({\rm mod}\;\;\; N)\\
&\omega_{k_1}+\omega_{k_2}+....+\omega_{k_S}=\omega_{k_{S+1}}+...+\omega_{k_M}.
\end{split}
\end{equation} 
The above processes convert $S$ waves into $T$ waves, where $S+T=M$.
A discussion of resonances  for the FPUT system  can be found in \cite{rink2001symmetry,rink2006proof,henrici2008results}. These studies deal with {\it four-wave resonances}, i.e. $M=4$, while in \cite{bustamante2019exact} the study is extended up to $M=6$. In \cite{bustamante2019exact}, it was proven that, for the FPUT system, for an arbitrary number of masses $N$, resonant processes converting 1 wave to $M-1$ waves or $M-1$ waves to 1 wave do not exist, for any $M\ne$2. Also, resonant processes converting 0 wave to $M$ waves, or $M$ waves to 0 wave do not exist, for any $M> 0$. This statement implies that {\it three-waves} resonant interactions (resonant processes converting 2 waves into 1 wave, or vice versa) are not allowed because this would require $\omega_{k_1+k_2} = \omega_{k_1}+ \omega_{k_2}$, which is not possible. Four-wave resonances for which two waves are converted into 2 waves are possible only for an even number of particles $N$; resonances are parameterised as follows:
\begin{equation}
\label{eq:4-wave_sol_k}
\{k_1,k_2; k_3,k_4\} = \left\{k_1,  \frac{N}{2} - k_1;  N-k_1, \frac{N}{2}+k_1\right\} \,, \qquad k_1 = 1, \ldots , \lfloor N/4 \rfloor,
\end{equation}
In general, each of these ``quartets'' produce energy exchange across the four modes involved, but there is no energy exchange across different quartets. This is somehow related to the existence of the integrable Birkhoff resonant normal form found in \cite{kappeler2009resonant} of order four. The case $S = T = 3$, corresponding to six-wave interactions, was discussed in \cite{onorato2015route} in the context of Wave Turbulence theory. Such solutions exist for any $N$ and wave numbers belonging to different resonances are connected. A detailed discussion of such solutions can be found  in \cite{bustamante2019exact}. To be mentioned is that solutions  for $M=5$ have been found if $N$ is divisible by 3 and $N>9$.

\subsection{Example of a formal perturbation expansion of the $\alpha$-FPUT problem}\label{subsec:pert_exp}
We consider the Hamiltonian in Eq. (\ref{eq:Ham_alpha}) for a system of $N$ masses, and use Eq. (\ref{action-angle}) to rewrite it in terms of action-angle variables $\{J_k,\theta_k\}$.
 The Hamiltonian reads
\begin{equation}
\begin{split}
&\frac{\mathcal H}{N}=\sum_k \omega_kJ_k+\epsilon \sum_{k_1,k_2,k_3} 
\tilde V_{1,2,3} \sqrt{J_1 J_2 J_3}\bigg[
\sin(\theta_1+\theta_2+\theta_3)\delta_{1,2,3}+\\
&+\sin(-\theta_1+\theta_2+\theta_3)\delta_{1}^{2,3}
\bigg],
\end{split} \label{eq:Ham_alpha_aa}
\end{equation}
with $\tilde V_{1,2,3}$ an appropriate interaction matrix that depends on $V_{1,2,3}$ in (\ref{eq:Ham_alpha}); we have introduced an $\epsilon\ll 1$ to indicate that the interaction Hamiltonian is a small perturbation. 
The Hamiltonian \eqref{eq:Ham_alpha_aa} is formally of the type in Eq. (\ref{eq:pert_kam}), yet the KAM theorem cannot be directly applied because the nondegenerate condition is not satisfied.
 Note that the perturbation Hamiltonian has two contributions: the first one converts 0 waves to 3 (and viceversa), the second one 1 to 2 (and viceversa). As previously mentioned, none of those processes are resonant.  The goal is then to perform a coordinate transformation from variable  $\{J_k,\theta_k\}$ to  $\{I_k,\psi_k\}$ to  ``push'' the phases to higher order in $\epsilon$. The procedure, indicated in \cite{arnold2007mathematical}, consists of seeking a generating function $S=S(\{I_k,\theta_k\})$ and the new Hamiltonian, $H(\{I_k\})$, as formal power series expansions:
\begin{equation}
\begin{split}\label{eq_can_transf_KAM}
&J_k=I_k+\epsilon \frac{\partial S}{\partial \theta_k},\;\;\;\;\; \theta_k=\psi_k+\epsilon \frac{\partial S}{\partial I_k}\\
&S(\{I_k,\theta_k\})=S_1(\{I_k,\theta_k\})+\epsilon S_2 \{I_k,\theta_k\}+...\\
& H (\{I_k)=H_0( \{I_k\}) +\epsilon H_1( \{I_k\})+...
\end{split}
\end{equation}
Exploiting the fact that $S_1, S_2,...$ are $2\pi$-periodic functions in $\{\theta_k\}$, one can use Fourier Series to solve at each order the  set of equations in (\ref{eq_can_transf_KAM}). Because of the absence of resonances up to order $\epsilon$,  the calculation can be  carried out in a straightforward way and the interesting result is that $H_1( \{I_k\})$ is identically 0.  However, despite this high gain, the transformation develops terms of order $\epsilon^2$, which are of the type as those included in the $\beta$-FPUT or in the DNLKG,  Eqs. (\ref{eq:Ham_beta}) or (\ref{eq:Ham_DNLKG}), but with different matrix elements. The new Hamiltonian at order $\epsilon^2$  reads as:
\begin{equation}
H (\{I_k,\psi_k\})=H_0 (\{I_k\})+\epsilon^2
\left(H_2^{0 \leftrightarrow 4}+
H_2^{1 \leftrightarrow 3}+
H_2^{2 \leftrightarrow 2}
\right),
\end{equation}
where the perturbation Hamiltonians $H_2^{S \leftrightarrow T}$, which convert $S$ waves into $T$ waves, are a function of $\{I_k,\psi_k\}$.
 As previously explained,   it has been shown in ~\cite{bustamante2019exact} that $0\leftrightarrow 4$ or $1\leftrightarrow 3$ processes are never resonant; therefore, those terms can be canonically eliminated at this order. 
 The $2\leftrightarrow 2$ contains both resonant and nonresonant term. The exclusion of all non resonant terms leads to an integrable Birkhoff normal form (resonant or not depending on the parity of $N$). In \cite{onorato2015route} it has been  conjectured that the recurrence behaviour of the $\alpha$-FPUT can be attributed to the normal form transformation. Recent analytical results and numerical computations, see \cite{ganapa2023quasiperiodicity}, have verified that, at least for small nonlinearity, the recurrence behaviour is well captured by the transformation that removes three-wave nonresonant interactions. 

The Wave Turbulence theory, being a statistical theory, is developed in the large box limit;
what appears to be a nonresonant (or quasi-resonant) interaction in $H_2^{2 \leftrightarrow 2}$ in a discrete system, may become a resonance in the large box limit \cite{aoki2006energy,onorato2015route,de2022anomalous}. Therefore, the standard procedure to describe the interaction of a large number of waves is not to start with the asymptotically integrable Birkhoff normal form, but with a perturbation Hamiltonian that converts 2 waves into 2 waves and contains also angles not related to resonances in the discrete system. 
In fact, quasi-resonances in the discrete system will become exact resonances in the large box limit; those exact resonances will play a major role in the transfer of energy between modes, and in the thermalization process.  Formally, the Hamiltonians in equations (\ref{eq:Ham_alpha}),  (\ref{eq:Ham_beta}),  (\ref{eq:Ham_DNLKG}), after the described change of coordinates,  can be rewritten in terms of the normal complex variables as:
\begin{equation}\label{eq:four_wave_ham}
H = \sum_{k=0}^{N-1} \omega_k |a_k|^2+ \epsilon^2\frac{1}{2}\sum_{k_1,k_2,k_3,k_4} T_{1,2,3,4} a_1^* a_2^* a_3 a_4  \delta^{1,2}_{3,4},
\end{equation}
where, despite the variable transformation, we have used the same letter $a$ for the sake of simplicity.
The matrix $T_{1,2,3,4}$ is different for each of the three models, see \cite{pistone2019universal}. The Hamiltonian in Eq. (\ref{eq:four_wave_ham}) is very general and applies to many systems. The only requirement is that the matrix element satisfies the following symmetries: 
\begin{equation}
	T_{1,2,3,4} = T_{2,1,3,4} = T_{1,2,4,3} =T_{3,4,1,2}\,.
\end{equation}
In the next Section, Wave Turbulence theory will be developed starting from the above Hamiltonian.

\section{Wave Turbulence Theory}\label{sec:WT}
The Wave Turbulence theory is a statistical theory that describes the spectral energy transfers in systems characterized by a large number of weakly interacting and dispersive waves.
In this Section, we will review the application of the Wave Turbulence theory to nonlinear particle chains. 
Relevant reviews focused on the general theory and applications can be found in 
\cite{zakharov2012kolmogorov, LVOVBOOK,nazarenko2011wave,galtier2022physics}.

 \subsection{Nonlinear chains in normal variables} We start from the general
Hamiltonian of a one dimensional chain of $N$ atoms with nearest neighbor anharmonic interaction potential defined in Eq.~\eqref{eq:four_wave_ham}. We recall that the wavenumber domain is spanned by $k\in \frac{2\pi}{L}{\mathbb Z_N}$ (integers modulo $N$), with $L=Na$. The equation of motion is given by the variation of the Hamiltonian 
\begin{equation}i \dot a_k = \frac{\partial {H}}{\partial a_k^*},
  \label{eq:ak-eq}
\end{equation}
and it reads
\begin{equation}
  i \frac{d a_k}{dt} = \omega_k a_k
  +\epsilon^2 \sum\limits_{123} T_{k,1,2,3} a_1^* a_2 a_3 \delta^{2,3}_{k,1}\,,
  \label{FourWaveEquation}
  \end{equation}
%
where the Kronecker delta is defined in \eqref{delta1,2+3},
and we use the short-hand notation $\sum_{12...}=\sum_{k_1,k_2,...}$. 
The sum over all modes includes then the {\it Umklapp} processes~\cite{ashcroft2022solid}.

\subsection{The wave action spectrum}
From the equation~\eqref{FourWaveEquation}, it is possible to derive an equation for the second moment of the statistics, i.e., the wave action spectrum. Such an equation is named the wave kinetic equation, and is at the core of Wave Turbulence theory. For weakly interacting waves, the wave kinetic equation plays a role analogous to the Boltzmann equation for a diluted ideal gas of particles.
Pioneering derivations of wave kinetic equations are found in~\cite{peierls1929,zakharov1966energy}.
The derivation may follow different paths, and we refer to the large literature and bibliography therein for the details and the many subtleties~\cite{zakharov2012kolmogorov,newell2011wave,nazarenko2011wave,eyink2012kinetic,chibbaro20184}.
We give here a formal sketch of the derivation in the case of four-wave interactions, underlying 
the most important steps. Rigorous mathematical derivations can be found in \cite{lukkarinen2011weakly,buckmaster2020kinetic,staffilani2021wave,deng2021derivation,deng2021full}; however, none of those results concerns particle chains directly. 

The wave-action spectrum for the discrete system, identified by~\eqref{FourWaveEquation}, in spatially homogeneous conditions, is defined by
\begin{equation}
n_{ k}(t) = \langle |a_{k(t)}|^2\rangle \,,
 \label{WaveAction}
\end{equation}
where $\langle\dots\rangle$ denotes averaging over a statistical ensemble to be defined below.
In the large-box limit, in which the wave numbers $\kappa=2\pi k/Na\in[0,2\pi/a]$ become a continuous variable, the meaningful finite quantity will be the wave-action spectral density (sometimes simply referred to as {\it action spectrum} for brevity), defined as
\begin{equation}
\tilde n_\kappa(t)=\lim_{L\rightarrow\infty}\frac{L}{2\pi} n_{ k}(t)\,.
 \label{WaveActionDensity}
\end{equation}
The wave kinetic equation describes the time evolution of this quantity.

\subsection{Derivation of the wave kinetic equation}
The derivation presented here follows mostly the one reported in the book of S. Nazarenko \cite{nazarenko2011wave}.
The starting point of the derivation is equation \eqref{FourWaveEquation}, which is also known as the Zakharov equation.
We first define a normalized frequency variable as
\begin{equation}
\Omega_k=\omega_k+\tilde \omega_{k}\,,\quad \text{with} \quad \tilde \omega_k= \epsilon^2 \big(2\sum_{k_1}T_{k,k_1,k,k_1}|a_{k_1}|^2-T_{k,k,k.k}|a_k|^2 \big)\,.
\end{equation}
We introduce a new variable ${ b_k}$ such that 
\begin{equation}
{b_k} = e^{i\Omega_{k}t}{ a_k}~,
\end{equation}
and rewrite the equation in the so-called {\it interaction representation}:
\begin{equation}
  i\dot b_k = 
  \epsilon^2 \sum\limits_{1 2 3} T_{k,1,2,3} b_1^\star b_2 b_3 e^{i\Omega^{k1}_{23}t} \delta^{k,1}_{2,3}\,,
  \label{FourWaveEquationOfMotionB}
  \end{equation}
  where $\Omega^{k1}_{23}\equiv \Omega_k + \Omega_1 - \Omega_2 - \Omega_3$. The parameter $\epsilon$ characterizes the strength of the nonlinearity.

The first crucial step to find an asymptotic statistical closure
is to consider that the dynamics is given by a weakly nonlinear evolution. 
For small nonlinearity, the linear time-scale $\tau_{\rm L}=2 \pi / \omega_k$ is much smaller than the nonlinear time-scale over which the spectral density will be shown to evolve, $\tau_{\rm NL}=O(\epsilon^4)$.
Thus, to filter out the fast linear oscillations, we introduce an auxiliary 
intermediate time $T$ such that $\tau_{\rm L}
\ll T \ll \tau_{\rm NL}$ and seek a solution at time $T$ as a perturbation
expansion in small $\epsilon$,
\begin{equation}\label{eq:pert-exp}
b_k(T)=b_k^{(0)}+\epsilon^2 b_k^{(1)}+\epsilon^4 b_k^{(2)} + {\rm h.o.t}\,.
\end{equation}
We then substitute this into (\ref{FourWaveEquationOfMotionB}) and solve order by order. At zeroth order, the result is time-independent:
$ b_k^{(0)}(T)=b_k(T=0)\label{definitionofa} $.
For ease of notation, in the remainder of the Section, the initial conditions are denoted by $b_k=b_k(T=0)$ when no explicit time dependence is indicated. 
The first iteration of (\ref{FourWaveEquationOfMotionB})
gives:
\begin{equation}
b_k \one(T) = - i \sum_{123} T_{k,1,2,3} 
 b_1^* b_2 b_3 \delta^{k,1}_{2,3} \Delta_T(\Omega^{k1}_{23}) \,, \label{FirstIterate}
\end{equation}
where $\Delta_T(y) = ({e^{iy T}-{1}})/{i T}$.
By iterating one more time, we obtain:
\begin{equation}
\begin{aligned}
b_l\two(T)=\sum_{123456}&\big( 
T_{k,1,2,3} T_{5,6,1,4} \delta^{k,1}_{2,3} \delta^{14}_{56}
 b_{2} b_3 b_4
 b_{5}^*  b_{6}^*
E(\Omega^{k56}_{234}, \Omega^{k1}_{23})\\
& - 
2 T_{k,1,2,3} T_{2,4,5,6}
\delta^{k1}_{23} \delta^{24}_{56}
 b_{1}^* b_4^* b_3 
 b_{5}  b_{6}
E(\Omega^{k14}_{356}, \Omega^{k1}_{23})\big) ~,
\end{aligned}
\label{SecondIterate}
\end{equation}
with $E_T(x,y)=\int_0^T \Delta_t(x-y)e^{i y t} d t $.
The next step consists in writing the evolution equation for
\begin{equation}
|b_k(T)|^2= |b_k^{(0)}|^2 + \epsilon^2 \big(b_k^{(1)}b_k^{(0)*} + c.c.\big) + \epsilon^4 \big( b_k^{(2)} b_k^{(0)*} + c.c. + |b_k^{(1)}|^2 \big)\,,
\label{eq:bks}
\end{equation}
using the
perturbative expansion terms
(\ref{FirstIterate})-(\ref{SecondIterate}). Now, we have to make some assumptions on the statistics of the wave field at initial time, in order to perform ensamble averaging. To characterize the distribution of the initial data, it is useful to think in terms of phases and amplitudes (or angles and actions), already introduced in~\eqref{action-angle}, i.e. $b_k^{(0)} = \sqrt{J_k}e^{-i \theta_k}$.
The assumption that turns out to be sufficient is that of a {\it Random Phase and Amplitude} (RPA) field~\cite{choi2005joint,nazarenko2011wave,X2}: the phases are  independent and identically distributed random variables, uniformly distributed over $[0,2\pi]$; the amplitudes are mutually independent random variables. The statistical closure enters in the phase averaging of  correlators of the wave field, thanks to the Wick's contraction rule~\cite{nazarenko2011wave}; for example for the fourth-order correlator it reads:
\begin{equation}
\label{eq:wick1}
	\langle  b_k^*b^*_1 b_2 b_3\rangle_{\theta,J} = \langle J_k\rangle_J \langle J_1\rangle_J \big(\delta^k_2\delta^1_3 + \delta^k_3\delta^1_2\big) + \big(\langle J_k^2\rangle_J -2 \langle J_1\rangle_J^2\big) \delta^k_1\delta^k_2\delta^k_3\,.
\end{equation}
Some important remarks are in order. First, the contraction rule is simply based on the fact that the argument of the complex exponential must be identically zero, or any contribution will vanish upon averaging over uniformly distributed angles. Second, notice that, so far, we have not invoked a specific distribution for the amplitudes. If we assume that the real and imaginary parts of the wave field $b_k$ are Gaussian-distributed, then the actions  $J_k$ are exponentially distributed. If this is the case, the second term of the r.h.s. of~\eqref{eq:wick1}, called the cumulant term, vanishes. In such case, the contraction rule coincides with a well known theorem for random Gaussian fields. However, this extra assumption of exponential amplitudes is not necessary for the derivation of the wave kinetic equation because the cumulant term has one extra delta function implying that, in the large-box limit, terms of this type become subleading ~\cite{nazarenko2011wave,eyink2012kinetic,chibbaro20184}. Instead, terms like the first one on the r.h.s. of ~\eqref{eq:wick1} allow for a direct closure in terms of wave-action spectrum, as we simply have $\langle J_k\rangle_J = n_k$.

By averaging over angles using Wick's contraction rule, we obtain:
\begin{equation}
\begin{aligned}
\langle b_k\one b_k^{*(0)}\rangle_\theta =& 
 - i \sum_{123} T_{k,1,2,3} 
\langle  b_k^*
 b^*_1 b_2 b_3\rangle_\theta
 \delta^{k,1}_{2,3} \Delta(\Omega^{k
 1}_{23}) + i \tilde \omega_k \langle J_k \rangle  T\,, \\
\langle  b_k^{*(0)} b_k\two\rangle_\theta=&
2\sum\limits_{123}
| T_{k,1,2,3} |^2 J_k
( J_2 J_3 - 2 J_1 J_2)
 \delta^{k,1}_{2,3} E(0,\Omega^{k1}_{23}),\\
 \langle | b_k^{*(1)}|^2\rangle_\theta=&
\sum\limits_{123}
|T_{k,1,2,3} |^2 J_1 J_2 J_3 \delta^{k,1}_{2,3}
|\Delta(\Omega^{k1}_{23})|^2.
\end{aligned}
\label{eq:bksq}
\end{equation}
As a consequence of the frequency renormalization, the first equation vanishes: the wavenumber pairings that would give a non-zero contribution after phase averaging in the triple summation are also contained in the term $i\tilde\omega \langle J_k \rangle T$, but with opposite sign. By plugging~\eqref{eq:bksq} into~\eqref{eq:bks}, and averaging over amplitudes, we obtain:
\begin{equation}
\label{eq:discrwke}
\begin{split}
&	n_k(T)-n_k(0) = \epsilon^4\sum_{123} |T_{k,1,2,3}|^2  \big( 2 n_k (n_2 n_3 - 2 n_1n_2) E(0,\Omega^{k1}_{23}) + \\
&+n_1 n_2 n_3 |\Delta(\Omega^{k1}_{23})|^2 \big) \delta^{k,1}_{2,3}\,.
\end{split}
\end{equation}
We are now ready to reach the wave-kinetic regime, which consists of a combination of the large-box limit ($L\to\infty$), followed by the small nonlinearity limit ($\epsilon\to0$)~\cite{nazarenko2011wave}. The order of the limits is a delicate issue, not unlike the proper limit scalings in the derivations of the Boltzmann equation~\cite{cercignani1972boltzmann}. If the $\epsilon\to0$ limit were taken first, the quasi-resonances would be ruled out of the picture, and a subsequent $L\to\infty$ limit would retain a substantially non-resonant character sometimes referred to as {\it frozen wave turbulence}. Recent derivations of the wave kinetic equation for PDEs have shown that one can take the two limits simultaneously with different possible scalings and successfully obtain the wave kinetic regime~\cite{buckmaster2021onset, deng2021propagation,dymov2021formal,dymov2021large,dymov2023formal}.

The large-box limit coincides with replacing sums with integrals and the Kronecker
deltas with Dirac deltas, as follows,
\be 
\left(\frac{2\pi}{L}\right)\sum_k \Longrightarrow \int_0^{2\pi/a} d\kappa ~, \quad
 \left(\frac{L}{2\pi}\right) \delta_{k,k'} \Longrightarrow \delta(\kappa-\kappa')\,,
\label{largeL} 
\ee
where $\kappa = 2\pi k/L$ for $L \rightarrow\infty$.
We then take the small-nonlinearity limit. 
Recalling that we are considering an intermediate time $T$, with $\tau_{\rm L}\ll T \ll \tau_{\rm NL}$, we can take for instance  $T\sim\frac{2\pi}{\epsilon^2 \omega_\kappa}$, so that $\lim_{\e\rightarrow0} T=\infty$. We have, as $T\to\infty$,
\begin{equation}
E_T(0,x)\sim (\pi
\delta(x)+iP({1}/{x})) T~~\text{and}~~|\Delta_T(x)|^2\sim2\pi T\delta(x)~,
\label{eq:wnlim}
\end{equation}
where $P(\cdot)$ indicates the principal value. 

Using~\eqref{largeL},~\eqref{eq:wnlim}, and~\eqref{WaveActionDensity}, Eq.~\eqref{eq:discrwke} yields:
\begin{equation}
\label{eq:contwke0}
\begin{split}
	\frac{\tilde n_\kappa(T)-\tilde n_\kappa(0)}{T \epsilon^4} & =  4\pi  \int_0^{2\pi/a}  d \kappa_{123} |T_{\kappa,1,2,3}|^2 \tilde n_\kappa \tilde n_1 \tilde n_2 \tilde n_3 \\
& \qquad\quad\quad \quad \times	\Big( \frac{1}{\tilde n_\kappa} + \frac{1}{\tilde n_1} - \frac{1}{\tilde n_2} - \frac{1}{\tilde n_3} \Big)  \delta^{\kappa,1}_{2,3} \delta(\Omega^{\kappa1}_{23}) \,,
\end{split}
\end{equation}
where the wave-number Dirac delta function, because of the periodicity of the Fourier space,  should be intended $({\rm mod }\;\;2\pi/a)$.
Let us define a new slow time variable $\tau = \epsilon^4 T$, and an action spectral density that is now function of $\tau$ instead of $T$: $\tilde n_\kappa(T) \to \tilde n_\kappa(\tau)$. We also drop all the tildes for the sake of notation. Since $\tau\to0$, the finite difference on the l.h.s. can be replaced by a time derivative. We finally obtain the wave kinetic equation
\begin{equation}
\label{eq:contwke}
\begin{aligned}
	\frac{\partial n_\kappa}{\partial \tau} = \mathcal I_{\kappa}= \eta_\kappa - \gamma_\kappa n_\kappa
\end{aligned}
\end{equation}
with
\begin{equation}\nonumber
\begin{aligned}
& \eta_\kappa =  4\pi  \int_0^{2\pi/a}  d  k_{123} |T_{\kappa,1,2,3}|^2   n_1   n_2   n_3\delta^{\kappa,1}_{2,3} \delta(\omega^{\kappa1}_{23})  \,,\\
	& \gamma_\kappa = 4\pi  \int_0^{2\pi/a} d k_{123} |T_{\kappa,1,2,3}|^2  [n_3(n_1+ n_2)-n_1n_2] \delta^{\kappa,1}_{2,3} \delta(\omega^{\kappa1}_{23}) \,.
\end{aligned}
\end{equation}
Strictly speaking, the above equation is valid at time $\tau=0$ (the right-hand side is computed at time $\tau=0$), although it is assumed that the equation is valid for larger times (the ``tildes'' have been removed from the $n_{\kappa}$). The ensemble average is  taken over initial data which are characterized by random phases and amplitudes; one of the major issues is related to proving that phases and amplitudes do not correlate for $\tau>0$. These properties have been shown to be preserved in time~\cite{eyink2012kinetic,chibbaro20184}, and stronger rigorous results have been obtained in \cite{deng2021propagation}
for the Nonlinear Schr\"odinger for dimensions $d \ge 3$. 
 
We remark that the time scale associated with the dynamics of the wave kinetic equation is of the order of $1/\epsilon^4$. Such result has been used  to estimate the thermalization time scale in the $\alpha$-, $\beta$-FPUT and in  DNLKG systems in the limit of a large number of particles, see  \cite{pistone2018universal}.

One may ask what happens if the Wave Turbulence machinery is used on an integrable system like the Toda lattice. The linear dispersion relation of the Toda lattice is identical to the $\alpha$- and $\beta$-FPUT; therefore, resonances exist; however, it turns out that the matrix elements $ T_{\kappa,1,2,3}$ computed on the resonant manifold  is identically zero  \cite{onorato2015route}; because of the integrability of the Toda lattice, it is expected that such behaviour holds at all orders.

\subsection{Collision Invariants, Entropy and Equilibrium distribution}\label{sec:coll-inv}
The wave kinetic equation in \eqref{eq:contwke0} has two conserved quantities: the total wave action, that is the number of waves or particles, and the energy:
\begin{equation} 
\begin{split}
&\mathcal N=\int_0^{2\pi/a} n_\kappa d\kappa,\;\;\;\; \mathcal E=\int_0^{2\pi/a} \omega_\kappa n_\kappa d\kappa.
\end{split}
\end{equation}
The conservation of energy is associated to the presence of the Dirac delta on frequencies. 
Unlike  the wave kinetic equation derived from partial differential equations (for example Nonlinear Schr\"odinger equation), because of the presence of Umklapp processes (the Dirac delta function is defined (mod $2\pi/a$), the momentum is not preserved for chains. The two conserved quantities have a microscopic counterpart; however, note that those quantities are only approximately preserved by the dynamics of the primitive equations of motion. Indeed, concerning the energy, the wave kinetic equation preserves only the energy associated with the unperturbed Hamiltonian; for what concerns the action, the microscopic equation \eqref{FourWaveEquation} preserves the total wave action; however, the original equations do not: it is only after the near identity transformations described in Sec. \ref{subsec:pert_exp} that the conservation in achieved. 

Just like for the Boltzmann equation for a gas of particles, for the wave kinetic equation it is possibile to define an entropy,
\begin{equation}\label{eq:entr_WT}
S(t)=\int_0^{2\pi/a} \ln n_{\kappa} d\kappa,
\end{equation}
that satisfies an $H$-theorem:
\begin{equation}
\frac{d S}{dt}\ge 0;
\end{equation}
details can be found in \cite{zakharov2012kolmogorov}. 
It is straightforward to show that entropy becomes constant if the wave action reaches the Rayleigh-Jeans distribution:
\begin{equation} \label{eq:RJ_WT}
n_\kappa^{(eq)}=\frac{T}{\omega_\kappa-\mu},
\end{equation}
where $T$ and $\mu$ are real quantities determined by initial conditions, playing the roles of the temperature and the chemical potential. Such quantities are associated to the conservation of  energy and wave action, respectively.
For $\mu=0$, \eqref{eq:RJ_WT}, the equipartition of energy is obtained, $\omega_\kappa n_\kappa=T=const$. Interestingly, the equilibrium distribution for the wave action obtained from the Wave Turbulence theory corresponds to the distribution obtained using arguments based on equilibrium statistical mechanics, see \eqref{eq:RJ_discrete}. Similarly,  the entropy defined in \eqref{eq:entr_WT} corresponds, once the equilibrium has been reached, to the one obtained in  equation \eqref{eq:entropy_discrete}.

\subsection{Equation for higher order moments and for the one-mode  probability density function }
Despite not directly used in the context of  FPUT chains, we complete this theoretical introduction to the Wave Turbulence theory by including a discussion on higher (than 2) order statistics. These ideas and theoretical results, reported in different frameworks, could be in principle tested on chains. 

A detailed description of the topic can be found  in the works~\cite{nazarenko2011wave,eyink2012kinetic,chibbaro20184}.
The approach to be presented, which traces back to the pioneering works of Peierls~\cite{peierls1929kinetischen} and Sagdeev~\cite{zaslavskii1967limits}, naturally includes the wave kinetic equation, and includes also higher-order statistics. The goal is to write an evolution equation for the single-mode  PDF (generalization to multi-mode PDF is also possible); the derivation is done through  the one-mode generating function $ Z(\lambda ,t)=\langle e^{\lambda J_k}\rangle, $ %
where $\lambda$ is a real parameter.  The PDF  can be written as an inverse Laplace
transform,
$
p(J_\kappa,t) = 
{1 \over 2 \pi i}  \int_{-i \infty}^{+i \infty} Z(\lambda, t)e^{-J_\kappa
\lambda}d\lambda. \label{distribution}
$
The one-mode moments are then given by
\begin{equation} 
M_\kappa^{(l)} \equiv \langle
J_\kappa^{l}\rangle 
= \int_{0}^{\infty} J_\kappa^l p(J_\kappa, t) \, dJ_k \label{momPDF}, 
\end{equation}
where $l \in {\mathbb N}$.
The perturbative expansion~\eqref{eq:pert-exp}, at $t=T$ is then used.
While at the zero-th order in $\e$ the equation
is closed, at the next orders, we must use the RPA statistical
hypothesis and perform the average using the Wick's contraction rule.
The large-box limit and the small-nonlinearity limit as in~\eqref{largeL}  and~\eqref{eq:wnlim} are then taken.
Finally, we perform amplitude averaging to obtain:
\begin{equation}\label{eq:PDF}
\dot  p = \frac{\partial}{\partial J_\kappa}\bigg[ J_\kappa \bigg( \eta_\kappa \frac{\partial p}{\partial J_\kappa} +\gamma_\kappa  p\bigg) \bigg]\,,
\end{equation}
where  $\eta_\kappa$ and $\gamma_\kappa$ are  given in~\eqref{eq:contwke}.
Using \eqref{momPDF}, one can get the evolution equation for the moments:
\begin{equation}
 \dot M^{(l)}_\kappa = -p \gamma_\kappa
M^{(l)}_\kappa + p^2 \eta_\kappa M^{(l-1)}_\kappa,\label{MainResultOne} 
\end{equation}
which, for $l=1$ is the standard kinetic equation~\eqref{eq:contwke}.
An important property of the one-mode PDF equation is the possibility to define an entropy functional and to show that this increases monotonically in time, and has a maximum precisely at the exponential distribution of $J_k$~\cite{nazarenko2011wave, eyink2012kinetic,chibbaro20184}:
\begin{equation}\label{eq:pdfeq}
p^{\rm(eq)}(J_k) = \frac{1}{ \langle J_\kappa \rangle} \exp\left(-\frac{J_\kappa }{ \langle J_\kappa \rangle}\right).
\end{equation}
The appropriate entropy functional writes
\begin{equation}\label{eq:H}
	S(p(J_\kappa)) = -\int dJ_\kappa \; p(J_\kappa) \ln\frac{p(J_\kappa)}{p^{\rm (eq)}(J_\kappa)}
\end{equation}
Notice that $S(p(J_\kappa))$ can be seen as a relative entropy with respect to the exponential distribution. 
It can be shown~\cite{eyink2012kinetic,chibbaro20184} that under the evolution of Eq.~\eqref{eq:PDF} one has $d S(p(J_\kappa)) / dt\ge 0$, with equality holding only at the equilibrium distribution, i.e.~$p^{\rm (eq)}$.
Interestingly, the stationary solution~\eqref{eq:pdfeq} of equation~\eqref{eq:PDF} for the one-mode PDF corresponds to the solution discussed in Eq.~\eqref{eq:sol_pdf}, obtained by making only considerations at equilibrium. 
Thus, rather than an initial requirement, a spontaneous approach to Gaussianity is retrieved as a consequence of the irreversible resonant dynamics. This has been tested numerically for the 2D problems of NLS or elastic plates~\cite{tanaka2013numerical,yokoyama2013weak,chibbaro2017wave,zhu2022testing}, and recently for the discrete Nonlinear Schr\"odinger equation ~\cite{onorato2020negative}.
Moreover, Eyink and Shi~\cite{eyink2012kinetic} have noted that the most general Boltzmann entropy coming from counting states can be written as
\begin{equation}
\begin{aligned}
	S_{tot} = & -\int d\kappa \int dJ_\kappa \; p(J_\kappa) \ln p(J_\kappa)\\
	=& \int d\kappa \; S(p(J_\kappa)) + \int d\kappa \; \ln n_\kappa + const \,,
\end{aligned}
\end{equation}
i.e. as the sum of two parts. The first part is the quantity defined in~\eqref{eq:H}, while the second part is the standard Wave Turbulence entropy~\eqref{eq:entr_WT}. Each of the two parts grows monotonically under the time evolution (i.e. satisfying its own ``$H$-theorem''), respectively, characterizing the spontaneous relaxation of the system towards the equilibrium amplitude PDF $p^{\rm (eq)}$ and the  equilibrium Rayleigh-Jeans distribution of the action, $n_\kappa^{\rm(eq)}$.

In summary, starting from a general Hamiltonian which can be written
as a linear harmonic main component perturbed by weakly nonlinear terms,
it is possible to write dynamical equations in terms of normal modes, and to
perform a perturbation analysis to calculate the statistical
correlators keeping only the terms at the leading order.  Yet, these
statistical equations are  not closed because of the nonlinearity as in
classical BBGKY hierarchy~\cite{balescu1997statistical}.  In order to close
them, we average over phases and amplitudes using the Random Phase and
Amplitude hypothesis, and we take the large-box limit and the
weak-nonlinearity limit. The stationary states towards which the system relaxes spontaneously correspond to the states obtained using classical statistical mechanics at equilibrium.

\newpage
\section{Finite-size effects}
\label{sec:NR}
In many systems and in numerical experiments, the number of sites $N$ is finite, and may be possibly small, as in the original experiments by Fermi and collaborators.
Moreover, the  coupling $\epsilon$, despite being small, is finite and the weak nonlinearity limit cannot be taken.
Both of these facts raise a question about the applicability of the Wave Turbulence framework to these experiments outside its rigorous asymptotic framework~\cite{kartashova2010nonlinear}.

In practice, we expect to distinguish two regimes in the FPUT dynamics.
One corresponds to the thermodynamic limit, even though the discrete chain is finite (but very large). Since $N$ is finite, the expected four-wave  exact resonances of the large-box limit do not necessarily fall onto the discrete resonant manifold. 
As put forward by Chirikov~\cite{chirikov1979universal}, the nonlinear interaction causes a broadening of the frequencies. When the broadening is such that there  is an overlap between separated frequencies,
the system becomes stochastic and resonances are allowed.
In FPUT, we therefore expect to retrieve the Wave Turbulence results when the broadening of the frequencies, related to the value of $\epsilon$, is large enough to make up for the discrete gaps (due to finite $L$, or $N$) that would prevent most resonances to exist. On the other hand, we remark that after taking the thermodynamic limit~\eqref{largeL}, it is clear that the small-nonlinearity limit~\eqref{eq:wnlim} involves the asymptotic collapse of the broadened distributions $\Delta_T$ and $E_T$ (i.e., near-resonances) onto one point in frequency space (i.e., exact resonance). Even though all of the ``mass'' of the delta function ends up being concentrated solely on the exact resonance point, this collapse does keep memory of all of the near-resonances, even in the $\epsilon\to0$ limit. Thus, also in the asymptotic regime, there is the same hidden delicate balance of the two joint limits, that combine in such a way that $L\to\infty$ and $\epsilon\to0$ must ``keep pace with each other''.

The second possible regime, when $N$ and/or $\epsilon$ are too small for the exact four-wave  resonances to be relevant, is qualitatively different.
First, it is necessary to remove the non-resonant contribution from the Hamiltonian flow, and check whether the next order of interactions allows for exact resonances that are contained in the resonant manifold.
It turns out that for any $N$ (even or odd), six wave interactions always exist for the FPUT systems;  the task of classifying such resonances was completed in \cite{bustamante2019exact}. 
To achieve this goal, two methods were used. The first method is based on the number
theory for the {cyclotomic polynomials}.
%
%
%
%
Specifically, the problem of finding exact resonances for FPUT system of
$N$ particles is equivalent to a Diophantine equation. These
calculations depend sensitively on the number of particles $N$, in
particular on the set of divisors of $N$. Another method that is
developed in \cite{bustamante2019exact} is the {\it pairing-off
  method}, where the set of wave numbers is paired to their ``buddy''
via the Umklapp scattering. 

Apart from  trivial resonances,
\begin{equation}
 k_1=k_4,\; k_2=k_5,\;k_3=k_6, \label{SixWaveTrivial}
\end{equation}
with all permutations of indices 4,5,6 (those are
responsible only for a nonlinear frequency shift), 
non trivial solutions that satisfy the equations
\begin{equation}
\begin{split}
&k_1+k_2+k_3-k_4-k_5-k_6=0\;\;\;\;\;\; ({\rm mod }\; N),
\\ 
&\omega_1+\omega_2+\omega_3-\omega_4-\omega_5-\omega_6=0,
\label{SIX}
\end{split}
\end{equation}
 for $k_i\in\mathbb{Z}$
exist for arbitrary $N$ for the FPUT systems. Symmetric resonances are found over the edge of the
Brillouin zone and are given by:
\begin{equation}
(k_1,k_2,k_3,-k_1,-k_2,-k_3),\label{SixWaveResonances1}
\end{equation}
 with $ k_1 + k_2 +k_3 = m N/2$ and
 $m = 0, \pm 1, \pm 2,...$ \, .\\
Quasi-symmetric resonances are found also over the
edge of the Brillouin zone, and  are characterized by one repeated wave
number:
\begin{equation}
(k_1,k_2,k_3,-k_1,-k_2,k_3), \label{SixWaveResonances2}
\end{equation}
with $k_1 + k_2 = m N/2$, $m= 0, \pm 1, \pm 2,...$ and all
permutations of the indices. 
These resonant sextuplets are interconnected, therefore it is expected that,
differently from discrete four-wave resonances, they can spread energy accross the spectrum.

One expects that the relevant dynamical equation takes the following form:
\begin{equation}
i \frac{d 
b_1}{d t}=  \omega_1 b_1+
\epsilon^2\sum_{234} T_{1,2,3,4}b_{2}^*b_3 b_{4}+\epsilon^4\sum_{23456} W_{1,2,3,4,5,6}b_2^*b_3^* b_4b_5 b_6 
\delta^{1,2,3}_{4,5,6}+O(\epsilon^5),
\label{evolub6} 
\end{equation}
where the sum at order $\epsilon^2$ should be made only on resonant terms and $N$ is selected in such a way that five-wave resonances are avoided.  Indeed, the four-wave
interactions cannot be removed because, even though they do not
contribute to the spreading of energy in the spectrum, they are
resonant and a canonical transformation suitable for removing those
modes would diverge.
The explicit form of the matrix elements $W_{1,2,3,4,5,6}$ has never been derived explicitly for FPUT, although it has been done for other systems, see for details
\cite{l2010spectrum,laurie2012one}. 

The final issue is now to understand whether a statistical description 
related to these resonances can be introduced.
Although from a mathematical point of view it seems hard to justify it for small $N$,  for $\epsilon$ small enough to avoid four-wave resonances triggered by a Chirikov mechanism, the existence of  a six-wave kinetic equation has been  conjectured for the FPUT chains, see \cite{onorato2015route}. The equation reads:
\begin{equation}
  \begin{split}
\frac{\partial n_\kappa}{\partial t}
  &= \epsilon^8\int_0^{2\pi/a}    d \kappa_{123456} |W_{\kappa,1,2,3,4,5}|^2
  n_k n_1 n_2 n_3 n_4 n_5 \times \\
 & \bigg[
    \frac{1}{n_\kappa}+\frac{1}{n_1}+\frac{1}{n_2}-
    \frac{1}{n_3}-\frac{1}{n_4}-\frac{1}{n_5}\bigg]{\delta}(\Delta \omega^{\kappa12}_{345})\delta^{\kappa,1,2}_{3,4,5}
\,.
  \label{SixWaveKEThermodynamic}
  \end{split}
\end{equation}
Note that the time scale associated to such dynamics is $1/\epsilon^8$, which is much larger than $1/\epsilon^4$ 
dynamics associated to the four-wave resonant kinetic equation.

\section{Thermalization in FPUT and DNKG models: numerical results}

Here, we focus  on the application of the WT theory to the analysis of the thermal equilibrium in chains, and to the prediction of  
the time-scale of such dynamics. Based on the kinetic description of waves
previously given, a nonlinear time scale, $T_{nlin}$, is associated
to resonant processes; consequently, it is possible to estimate the
time scale associated with the equipartition, $T_{nlin}\sim T_{eq}$.  
Specifically, the time-scale is related 
to the first nonlinear term which admits non-trivial resonances in the kinetic equation. By construction, the time-scale will be inversely proportional to a power of  $\epsilon$, the
small coupling parameter of the system. As mentioned, 
for a four-wave process  
$T_{nlin}\sim \epsilon ^{-4}$, and for the
six-wave process 
$T_{nlin}\sim \epsilon ^{-8}$.

In order to
establish the time of thermalization, 
it is customary to consider the following metric
entropy~\cite{livi1985equipartition}:
\begin{equation}\label{eq:entr_livi}
s(t)={-}\sum\limits_{k} f_k \log f_k\;\; {\rm with}\;\;\; f_k = \frac{N-1}{\sum_j \omega_j \langle |a_j(t)|^2\rangle}\omega_k \langle |a_k(t)|^2 \rangle,\;\;\; 
\end{equation}
where $\langle...\rangle$ defines the average over the realizations characterized by the same initial spectrum but different random phases. Such entropy, although not the one defined in (\ref{eq:entr_WT}), shows, at least for the numerical simulations performed, see for example  \cite{onorato2015route}, a monotonic behaviour. The entropy in eq. (\ref{eq:entr_livi}), identically zero  when the equipartition of energy has been reached, has been used in numerical computation to establish  the time at which thermalization takes place.

\subsection{$\alpha$- and $\beta$-FPUT chains}

\subsubsection{Small $N$ regime}
\begin{figure}
\includegraphics[width=0.5\linewidth]{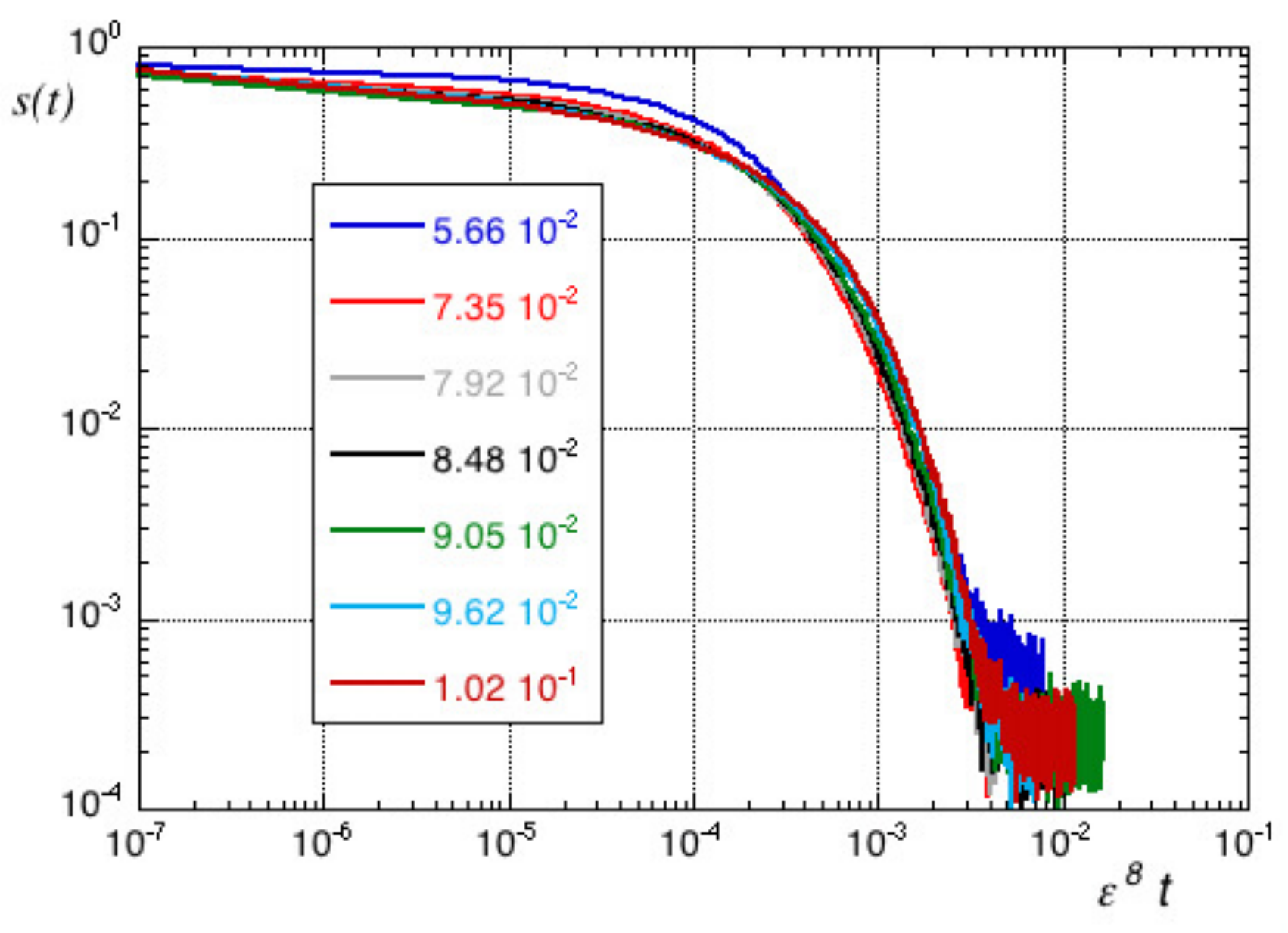}
\includegraphics[width=0.425\linewidth]{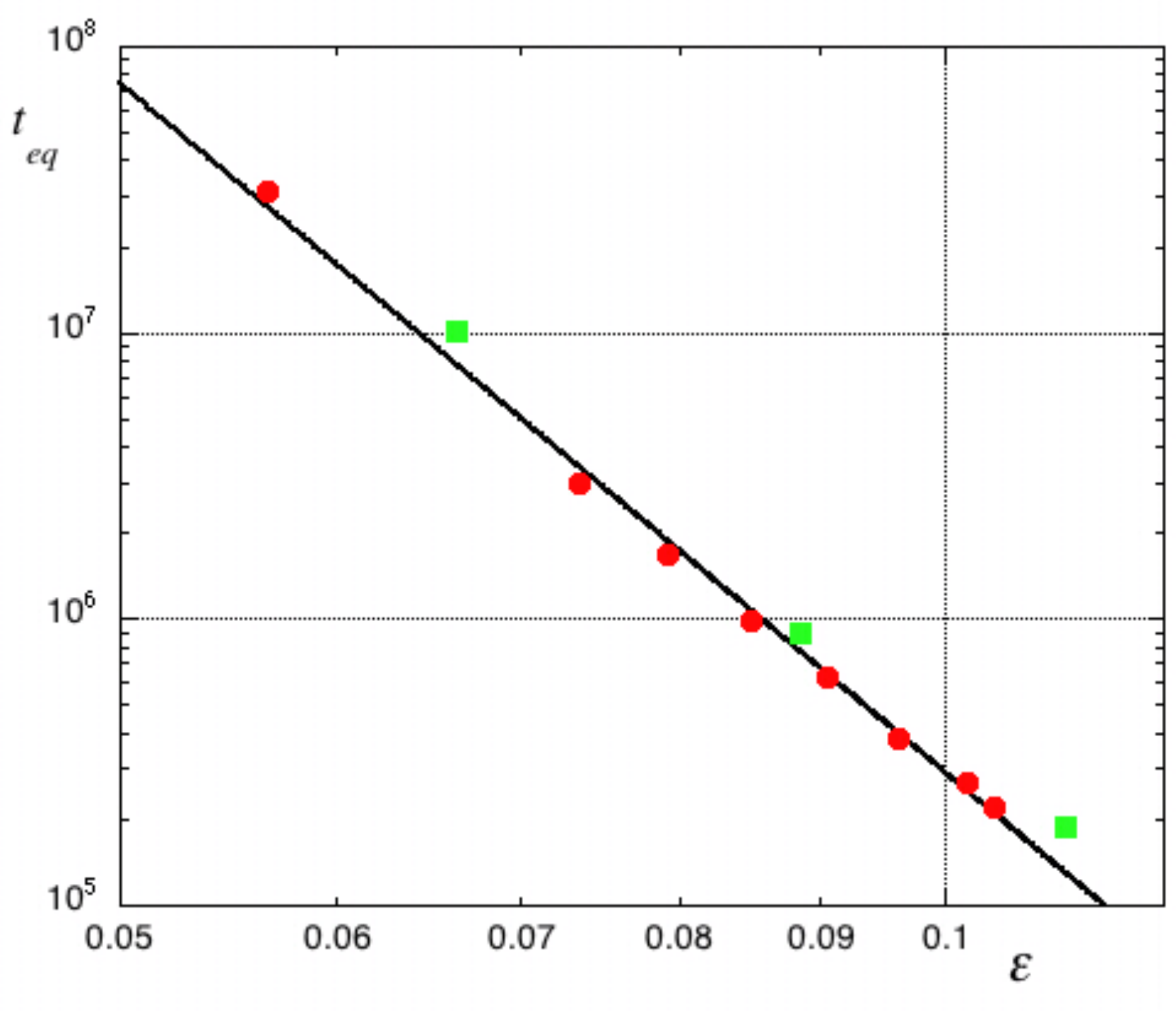}
\caption{Figure taken from~\cite{onorato2015route}. Left: Entropy $s(t)$ as a function of $\epsilon^8 t$ for different
  simulations  of the $\alpha$-FPU system 
  characterized by different values of $\epsilon$. Right: Equilibrium time as a function of $\epsilon$ in
  Log-Log coordinates. See  ~\cite{onorato2015route}  for details.
   }
 \label{entropy_coll}
\end{figure}
 In the original report by Fermi and collaborators, due to the limited amount of computational resources, simulations were performed with $N$=16, 32 and 64. Nowadays, we know that, although the number of masses is considered small for a rigorous application of statistical mechanics, numerical computations of the primitive equations of motion give evidence, at least for the initial data considered, of equipartition of energy among the normal modes. For the $\alpha$-FPUT and in \cite{lvov2018double} for the $\beta$-FPUT, it has been conjectured  in \cite{onorato2015route} that for such low number of masses, a six-wave kinetic equation should rule the slow dynamics to equipartition. 
Therefore, the characteristic evolution time
 of the spectral energy density in both models 
should be proportional to $1/\epsilon^8$.

\begin{figure}[h]
\begin{center}
\includegraphics[width=0.5\linewidth]{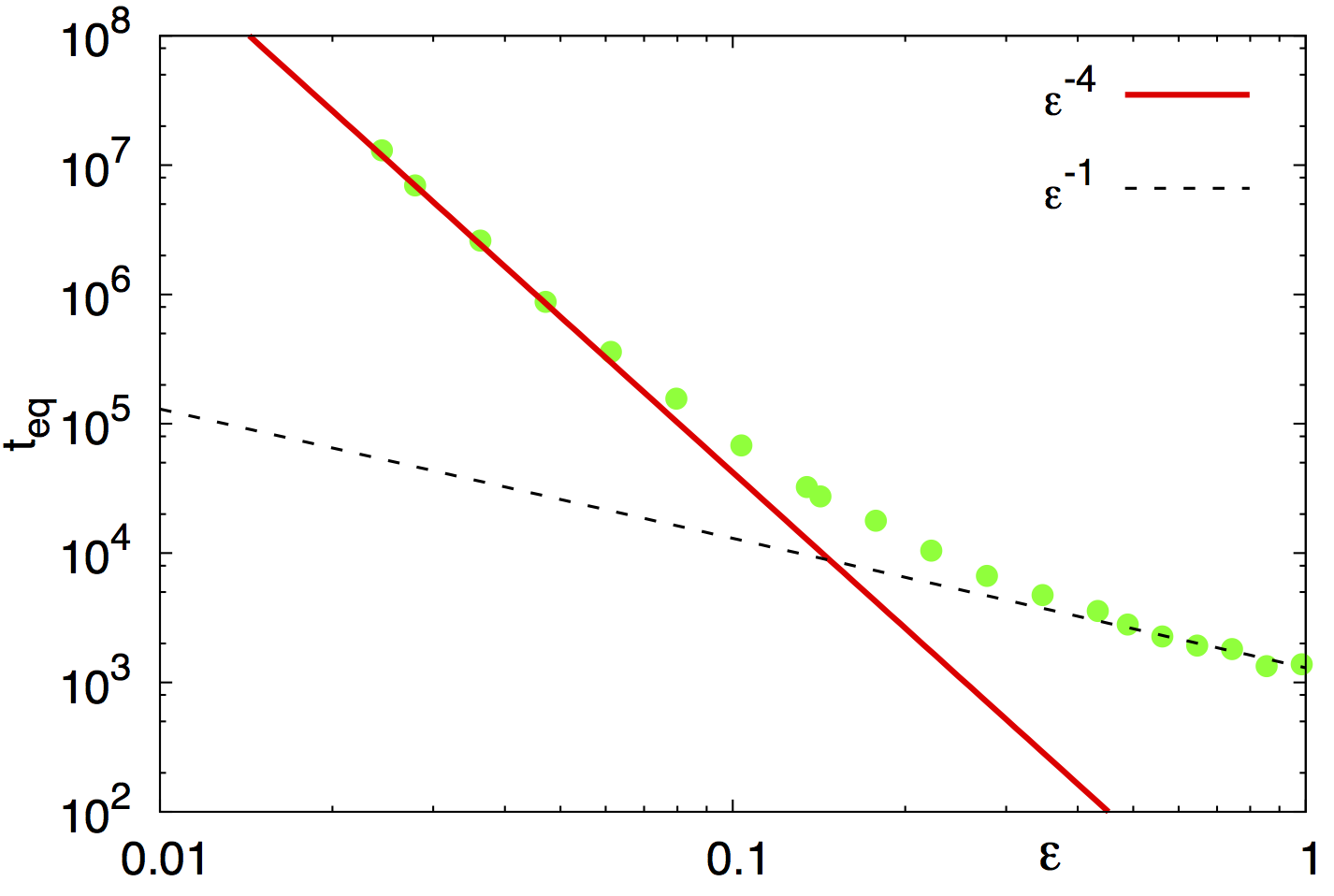}
\caption{Figure taken from~\cite{lvov2018double}. Equilibrium time $t_{eq}$ as a function of $\epsilon$ in
  Log-Log coordinates for the $\beta$-FPUT with $N$=32.  Dots represent numerical experiments. The re
  straight line corresponds to power law of the type $1/\epsilon^{4}$ (in the notation used in the text, it corresponds to
   $1/\epsilon^{8}$), while the dashed line to $1/\epsilon$ ($1/\epsilon^2$ in our notation); see \cite{lvov2018double} for details.
  } \label{fig:scaling}
\end{center}
\end{figure}   
In Figure \ref{entropy_coll} the major results found in \cite{onorato2015route} are reported: on the left the entropy defined   in (\ref{eq:entr_livi}) is displayed as a function of time, scaled with $\epsilon^8$ for different values of the small parameter $\epsilon$. Interestingly, all entropies collapse into a single curve, highlighting that the 1/$\epsilon^8$ is the correct scaling. On the right, the equipartition time scale as a function of $\epsilon$ is shown: again a clear power law $\sim \epsilon^{-8}$ is observable, see \cite{onorato2015route}  for details on initial conditions. Similar results can be obtained for the $\beta$-FPUT, see Figure \ref{fig:scaling}, where the time of equipartition is shown against the small parameter $\epsilon$.
Note that the definition of $\epsilon$ in \cite{lvov2018double} corresponds to $\epsilon^2$ in our notation; therefore, the $1/\epsilon^4$ displayed in the figure corresponds to  $1/\epsilon^8$ in the notation adopted in the present review. 
More simulations performed with different initial conditions on $\alpha$- and $\beta$-FPUT with $N=64$ can be found in 	
\cite{pistone2019universal}. The results support the conjecture that there exists a regime in which six-wave interactions are responsible for the 
equipartition of energy among the degrees of freedom.
At larger enough nonlinearity (for fixed $N$), a clear transition toward a different scaling, compatible with $1/\epsilon^{-1}$ ($\epsilon^{-2}$ in our notation), is visible. Its origin is unclear, although the Chirikov frequency overlap criterion~\cite{chirikov1979universal} has been considered in \cite{lvov2018double} for explaining  the transition between the regimes $1/\epsilon^8$  and $1/\epsilon^2$. The Wave Turbulence approach allows to estimate both the frequency shift due to nonlinearity and the broadening of frequencies, allowing for interactions also in discrete conditions, see \cite{lvov2018double}. 


\subsubsection{Large N regime}

We have mentioned that in the large $N$ limit (keeping the spacing, $a$, between masses constant) four-wave exact resonances exist, both for $\alpha$- and $\beta$-FPUT, and
  the standard four-wave kinetic equation can be recovered (see equation \ref{eq:contwke}). 
 The time scale associated with such an equation is $T_{nlin}\sim 1/\epsilon^4$; therefore,  
 one expects to observe the equipartition of energy on such time scale. 
 Numerical results~\cite{pistone2018universal},  shown in Figure \ref{fig:alpha1024},
provide a reasonable agreement between numerics and   the theoretical predictions for both $\alpha$- and $\beta$-FPUT models (in the figure $\epsilon_{\alpha}$ and $\epsilon_{\beta}$ are proportional to $\epsilon^2$). The exponent in the $\alpha$-FPUT is  steeper than predicted, probably because the thermodynamic limit has not been yet reached numerically.  Yet, as already explained in Section \ref{sec:NR}, in all numerical experiments $N$ is finite, thus finite-size effects might play a role.  Definitely, such issues should be investigated with further computations. 
Moreover, as it is well known, the $\alpha-$model suffers from the problem of unbounded trajectories at high-energy, and hence only very weak-nonlinearity may be explored, for which even larger $N$ are needed.
\begin{figure}
\includegraphics[width=0.5 \linewidth]{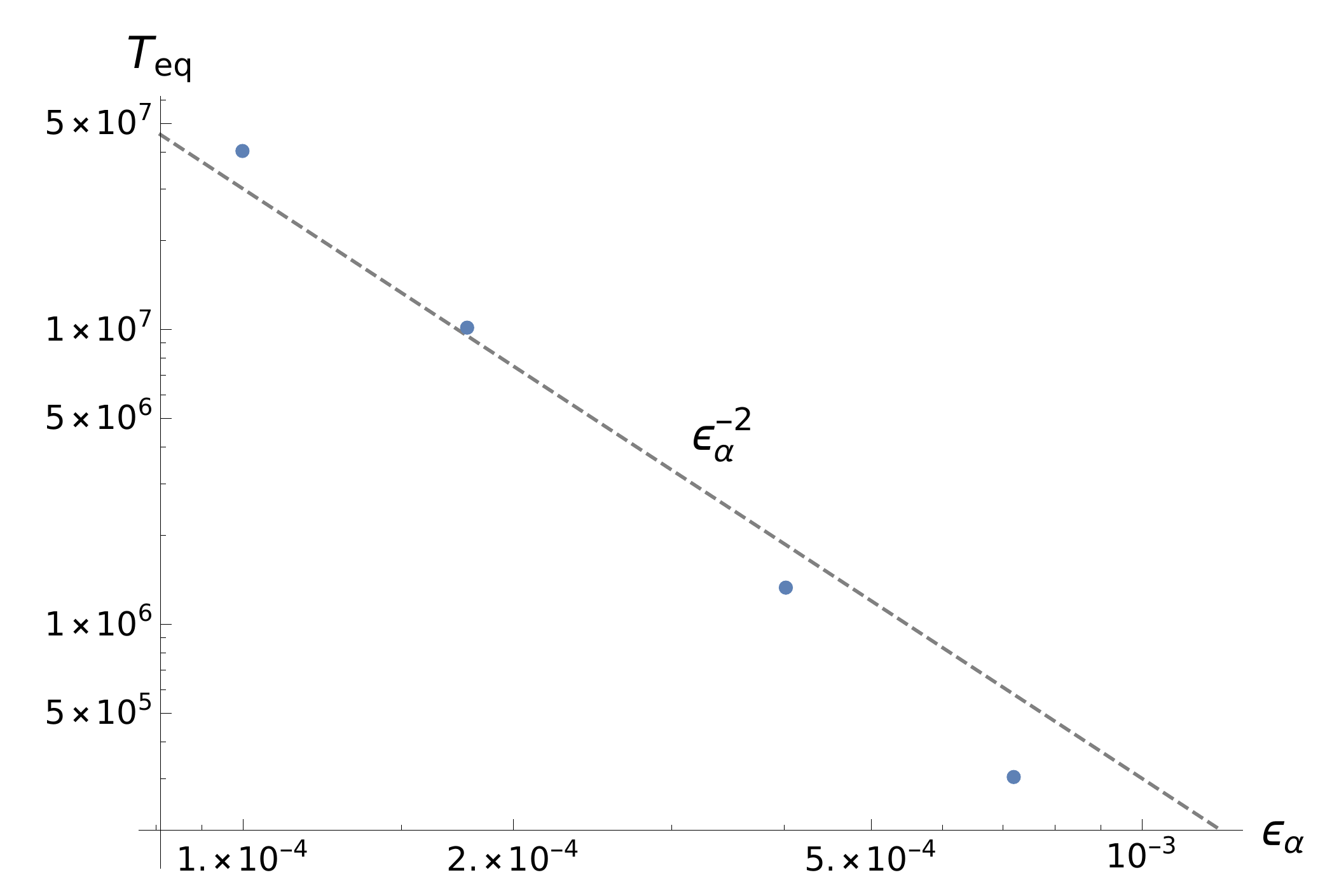}
\includegraphics[width=0.5 \linewidth]{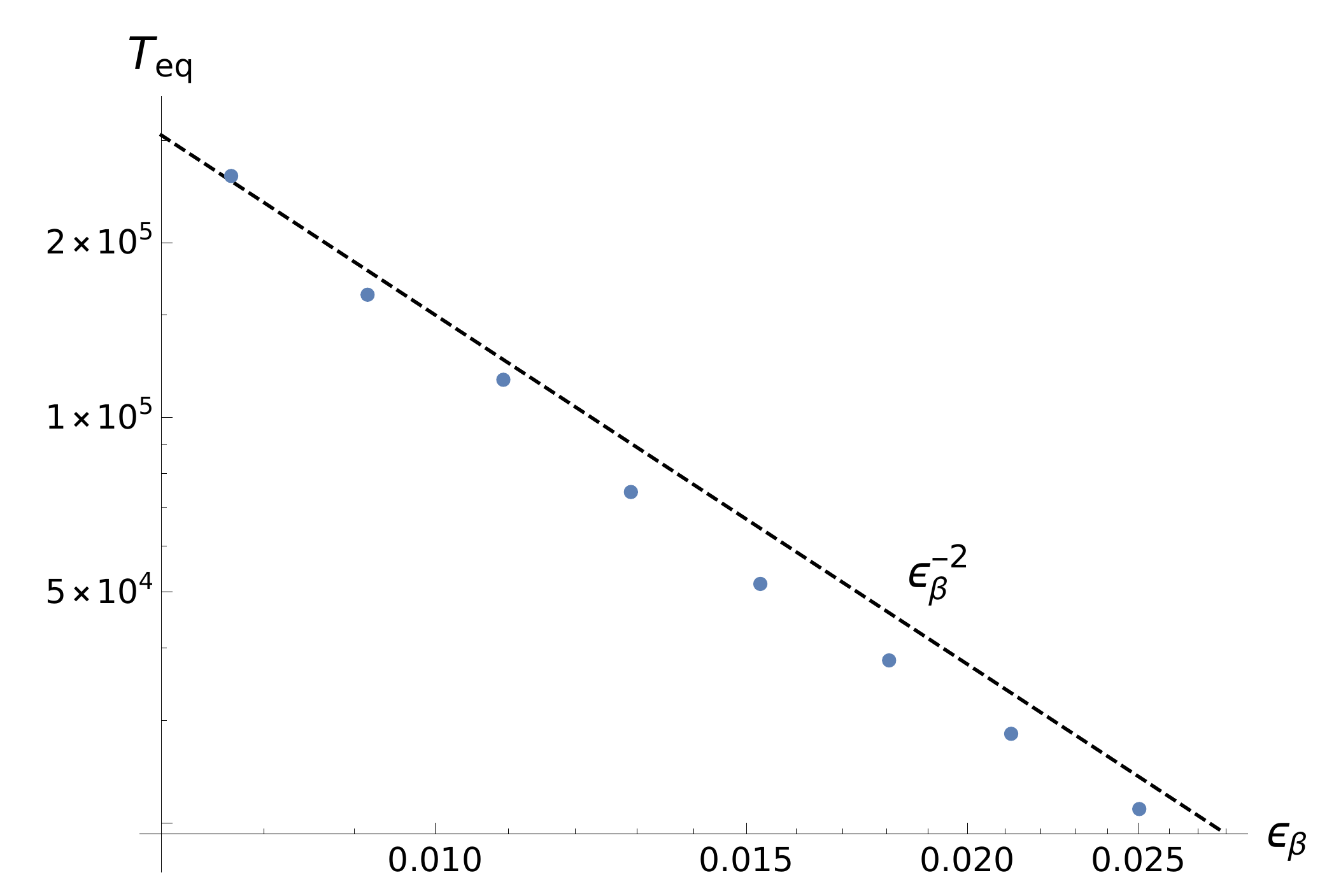}
\caption{Equipartition time scale for the (a) $\alpha$-FPUT and  (b)$\beta$-FPUT with $N=1024$.
}
\label{fig:alpha1024}
\end{figure}

\subsection{Verification of the Wave Turbulence expectations for other models}
In a recent work~\cite{fu2019universal}, the $FPUT$ model has been  generalised as follows:
\begin{equation}
{\mathcal H}_=\sum_{j=0}^{N}\left(\frac{1}{2 }p_j^2+\frac{1}{2}(q_{j+1}-q_j)^2+\frac{\lambda \epsilon^{-\frac{(n-2)}{2}}}{n}(q_{j+1}-q_j)^n\right).
\end{equation}
Assuming that Wave Turbulence applies in the thermodynamic limit, one may expect that the  $T_{nlin}\propto \lambda^{-2}\epsilon^{-(n-2)}$ for $n\geq 4$.
This finding has been corroborated by extensive numerical simulations~\cite{fu2019universal}, as  shown in Figure \ref{fig:fu}.
The deviation decreases with  increasing  $n$. This deviation is attributed to the extra vibration modes excited by the asymmetric potential ~\cite{fu2019universal}.
The results seem to be robust to changes in the mechanism of breaking integrability~\cite{fu2019universalT,fu2019nonintegrability}.
\begin{figure}
\centering
\includegraphics[scale=1]{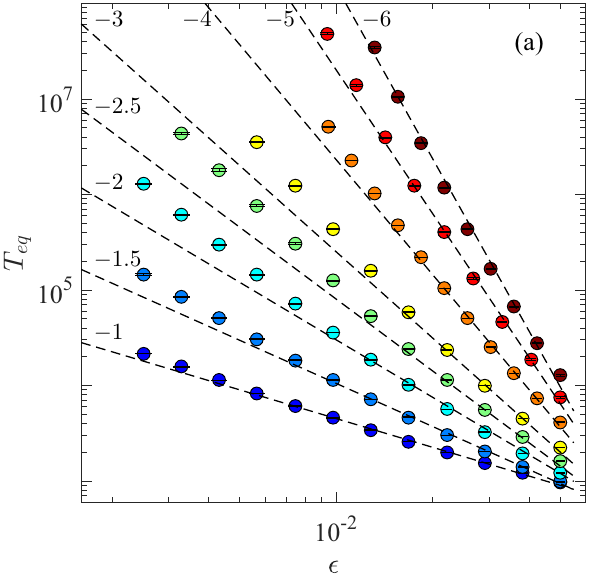}
\caption{Figure taken from ~\cite{fu2019universal}. The equipartition time $T_{eq}$ as a function of energy density $\epsilon$ in Log-Log scale, for symmetric models $V (x) = |x|^d /d$ with $d = 3, 7/2, 4, 9/2, 5, 6, 7, 8$ from bottom to top, and dashed lines with slope $2-d$ are drawn for reference. All simulations are performed for $N = 2048$. }
\label{fig:fu}
\end{figure}


The Wave Turbulence predictions have been extensively studied on the  Discrete Nonlinear Klein-Gordon (DNKG), or $\phi^4$, chain, see  Eq. (\ref{eq:KG}).
The DNKG has been  studied in relation to the breathers \cite{gorbach2003discrete,koukouloyannis2009stability,flach2008discrete}, and also to equilibration dynamics \cite{de2016equilibration,danieli2019dynamical}.
With a particular regard to the thermalization problem, the model has allowed for some interesting analytical and semi-analytical developments \cite{parisi1997approach,ponno2000analytical,giorgilli2015extensive}.
The $\phi^4$ chain was considered in an original way to shed light on the problem of thermalization in a pioneering study  \cite{fucito1982approach}. 
In this work, the concept of a metastable regime was put forward, observing that the system approaches equilibrium but only with a very slow dependence on time, so that the transient non-equilibrium state may be mistakenly taken as a steady one, if the observation time is too short.

The first application of Wave Turbulence theory to DNKG was done in \cite{pistone2018thermalization}. Assuming periodic boundary conditions, the dispersion relation for the DNKG equation, reported in equation (\ref{eq:disp_rel_DNLKG}), admits for any value of $s$ the same kind of resonances as those for the FPUT problem: for $N$ odd, four-wave resonant interactions do not exist; for $N$ even, four-wave resonances exist, however, they are of Umklapp type and the quartets are isolated; six-wave resonant interactions exist for $N$ even or odd. In the thermodynamic limit, four-wave resonances are connected and a four-wave kinetic equation can be derived.
A campaign of simulations has been carried out with different values of $N$  to test the theoretical predictions \cite{pistone2018thermalization}.
\begin{figure}[h]
\begin{center}
\includegraphics[width=0.75\columnwidth]{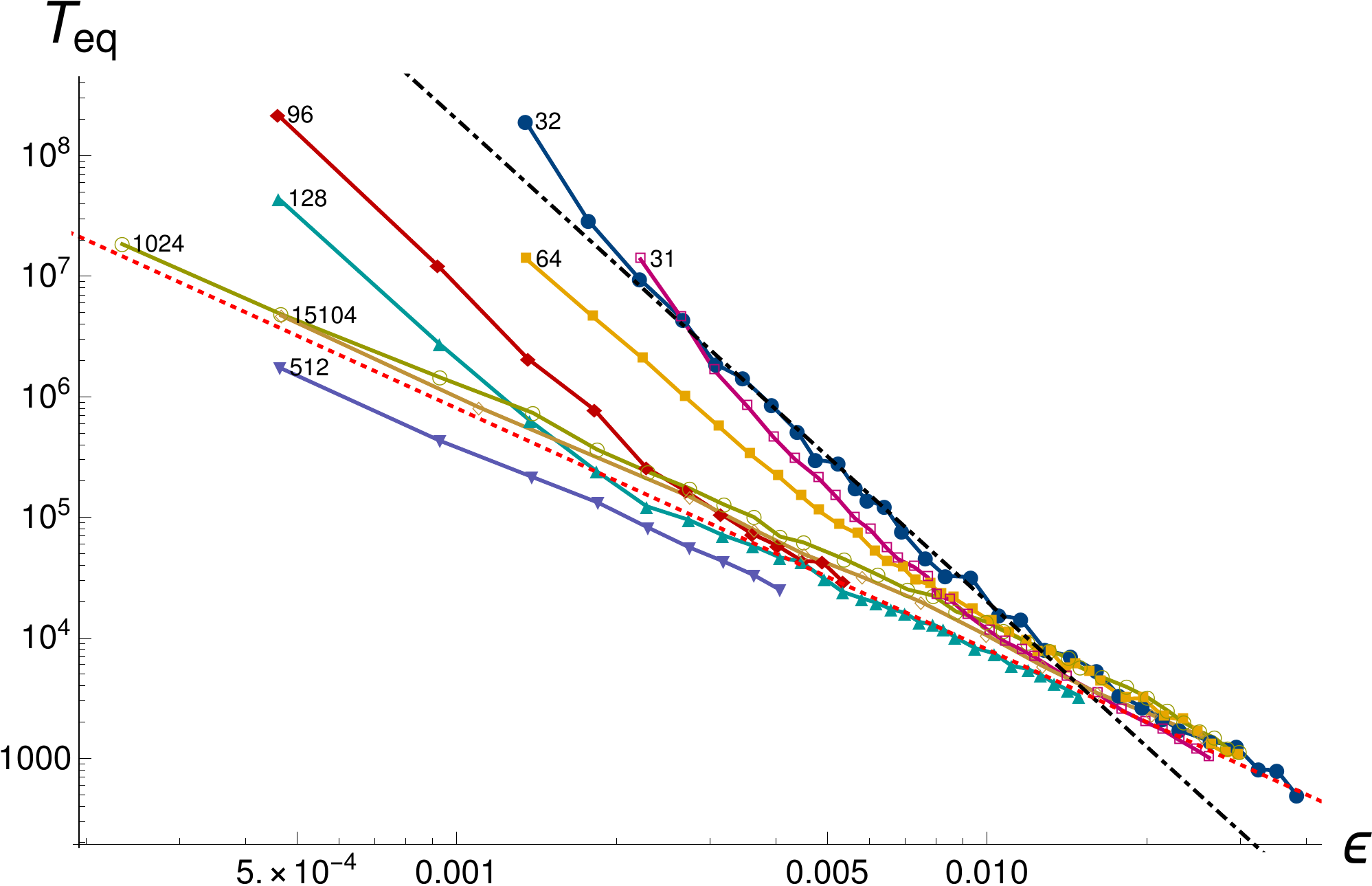}
\caption{
Figure taken from \cite{pistone2018thermalization}. The scaling of $T_{eq}$ as a function of $\epsilon$ for different values of $N$, with $s=1$ for the DNKG system.  Scaling laws $\epsilon^{-2}$ and $\epsilon^{-4}$ in red dotted and black dash-dotted lines.
}
\label{fig.condition}
\end{center}
\end{figure}
In Fig.~\ref{fig.condition}, the behaviour at varying initial $\epsilon$ and the dependence of $T_{eq}$ on $N$ are shown, as obtained in~\cite{pistone2018thermalization}.
The transition between two scalings, $T_{eq}\sim \epsilon^{-2}$ and $T_{eq}\sim \epsilon^{-4}$ is  well visible, ($\epsilon^{-4}$ and $\epsilon^{-8}$ in our notation).
For larger $N$ the transition turns out to manifest increasingly later until it disappears for $N\ge 512$, meaning that in the thermodynamic limit 
$T_{eq}\sim \epsilon^{-2}$, as predicted by the wave kinetic equation.
When $N$ is small the discrete scaling $T_{eq}\sim \epsilon^{-4}$ ($\epsilon^{-8}$ in our notation), related to six-wave resonances, is retrieved.

The Discrete Nonlinear Schr\"odinger equation has also been investigated in terms of the Wave Turbulence theory. Despite the time scale of equipartition has not been considered, the probability density functions of wave action at fixed wave number has been computed, showing a good agreement with the predicted exponential curve, see (\ref{eq:pdfeq}). Moreover, in \cite{onorato2022equilibrium} it has been shown  that negative temperature states corresponds to stationary solutions of the wave kinetic equation. The issue of negative temperatures is beyond the present review and the reader is invited to refer to the recent review \cite{baldovin2021statistical}.


\section{Anomalous Correlators\label{AnomalousCorrelators}}
In the Wave Turbulence theory, the standard object
to look at is the second-order correlator, $\langle
a_k(t)a^*_l(t)\rangle$, 
which, in the homogeneous case, is related to the wave action
spectral density function, i.e. the {\it wave spectrum}, $n_k=n(k,t)$.
However, one should note that the complex normal variable, as defined
in Wave Turbulence, is a complex function also in
physical space. Therefore, the second-order statistics are not fully
determined by the above correlator.  The so called ``anomalous
correlator'', $\langle a_k(t)a_l(t)\rangle$, may be also relevant~\cite{l2012wave,zakharov1975spin}. Those correlators first appeared in
the well known Bardeen, Cooper, and Schriffer (BCS) theory of
superconductivity \cite{BCSTheory}.  Subsequently, they have been studied in S--theory.
It is worth noting that in the S--theory \cite{zakharov1975spin,l2012wave} 
   the existence of anomalous correlators was connected with
  coherent pumping in the system, with the anomalous correlator being
  a measure of partial coherence for exiting waves. 
  Anomalous correlations were shown to play an important role
in explaining numerical observations of nondecaying oscillations
around a steady state in a turbulence--condensate system modeled by
the Nonlinear Schr\"odinger equation
\cite{NLSPhaseCorr,Dyachenko:92,NLSsim2}. Such oscillations, corresponding
to a fraction of the wave action being periodically converted from the
condensate to the turbulent part of the spectrum, were shown to be
directly due to phase coherence \cite{NLSPhaseCorr}.  In
\cite{guasoni2017incoherent}, a system of Coupled Nonlinear
Schr\"odinger equations has been considered, and specific attention was
focused on the phenomena of recurrence of incoherent waves observed
in the early stages of the dynamics. The authors derived a variant of
the wave kinetic equation which includes anomalous correlators; the
peculiarity of such an equation is that it is capable of describing
properly the recurrence phenomena observed in the simulations.

Under
the hypothesis of homogeneity, the anomalous correlator  is related to the anomalous
  spectrum, $m_k=\langle a_ka_{-k}\rangle$~\cite{zaleski2020anomalous}. If phases were random, this quantity would be zero, yet in the nonlinear evolution of some systems, this is not the case. One of the main tools used to derive the theory is the Wick's
contraction rule~\cite{nazarenko2011wave}, that allows one to split higher order correlators as
a sum of products of second order correlators, plus cumulants.  To
explain analytically the existence of the anomalous correlators, it is
necessary to use the more general form of the Wick's rule.
Following    \cite{LVOVBOOK}, the following expressions can be written:
\begin{equation}
\begin{split}
&\langle a_k^* a^*_l a_p a_n \rangle
= n_k n_l(\delta^k_p\delta ^l_n + \delta^k_n\delta ^l_p) +
m^*_k m_p\delta_{kl}  \delta _{pn}, \\
&\langle   a_k^* a_l a_p a_n\rangle =
n_k m_p(\delta_k^l \delta _{pn}+ \delta_k^n \delta_{lp})+ n_k \delta^p_k m_l\delta_{nl},\\
&\langle   a_k a_l a_p a_n\rangle = m_k m_l (\delta_{kp}\delta_{ln}+\delta_{kl}\delta_{pn}+\delta_{kl}\delta_{pn}).
\label{eq:wick}
\end{split}
\end{equation}
With the aim of studying the presence of anomlaous correlators in the $\beta$-FPUT, numerical simulations with  periodic boundary conditions have been considered in ~\cite{zaleski2020anomalous}.  
The results give evidence of the existence of such correlators also in particle chains;
the development of such a regime
corresponds to a tendency of the system to develop standing waves in
the original displacement variable $q_j(t)$.


\section{Thermal transport in FPUT chains}
\label{sec:12}

\subsection{Thermal conductivity in a 1D crystal}

This classical problem considers a 1D chain of particles interacting via nearest-neighbour potential, see Eq.~\eqref{eq:6.1}, see \cite{lepri2003thermal,lepri2016thermal,benenti2020anomalous}. The length of the chain is given by $L=aN$, where $a$ is the inter-particle distance at rest (from now on $a=1$),
and the thermodynamic limit $L\to\infty$ is taken as described in Section~\ref{sec:WT}. 
The Fourier space is spanned by the variable $k\in [-\pi,\pi]$. We conventionally assume the two ends of the chain to be in contact with two thermostats at temperature $T_+$ and $T_-$ with $T_+>T_-$, at the left and right ends, respectively. According to the second law of thermodynamics, the stationary state will be characterized by a net, spatially homogeneous current of thermal energy flowing from the hot to the cold thermostat. The thermal conduction properties of the system are intended as the relations that govern this flux of thermal energy, $J_e$, as a function of the system's macroscopic properties, such as $L$ , $\bar T = (T_+-T_-)/2$, and $\Delta T = T_+-T_-$. A fundamental quantity capturing the thermal conduction behavior is the thermal conductivity, defined as
\begin{equation}\label{eq:12-1}
	\kappa_e = J_e \frac{L}{\Delta T},
\end{equation}
i.e., the energy current per lenght-normalized temperature difference. 
The thermal energy is transported by wave-like excitations (or phonons), the eigenmodes of the linearized system, whose envelope propagates at speed $v_k=d\omega_k/dk$~\cite{lepri2003thermal,spohn2006phonon,dhar2008heat}. 
At any fixed point in the bulk of the chain, the net energy current is thus given by
\begin{equation}\label{eq:J12}
	J_e=\int _{-\pi}^\pi j_k \, dk,\quad\text{with}\quad j_k =\omega_k v_k n_k= (e^+_k - e^-_k)|v_k|\,,
\end{equation}
where $e_k=\omega_k n_k$, with the superscripts $+$ and $-$   indicate the restrictions  to $k\in[0,\pi]$ and $k\in[-\pi,0]$, respectively. 
Plugging this into~\eqref{eq:12-1}, we obtain  a general formula for the thermal conductivity,
\begin{equation}\label{eq:12-3}
	\kappa_e = \frac{L}{\Delta T}\int _{-\pi}^\pi j_k\, dk.
\end{equation}

\subsubsection{Harmonic chain and ballistic conduction}
In the case of harmonic oscillations,
the wave modes are mutually independent and propagate unperturbed through the system. In the stationary state, the phonons emitted by the thermostats at the two ends have average temperature $T_+$ and $T_-$, respectively. 
Because of the geometry of the system, 
in proximity of the left and right ends the modes $k\in[0,\pi]$ and $k\in[-\pi,0]$ are equipartitioned with energy density $e^+_k=c_L T^+/(2\pi)$ and $e^-_k=c_L T^-/(2\pi)$, where $c_L$ is the specific heat capacity per unit length~\cite{lepri2003thermal}.
 In stationary conditions, these expressions will also hold for any point in the bulk of the chain and therefore, substituting in Eq.~\eqref{eq:12-3}, the thermal conductivity reduces to
\begin{equation}\label{eq:12-4}
	\kappa_e \simeq \frac{L}{2\pi}\int _{-\pi}^\pi c_L |v_k| \,dk.
\end{equation}
Note that this is an oversimplification coming from implicitly assuming a mode-independent {\it transmission coefficient} of $1$ between the thermostats and the chain~\cite{dhar2008heat}. However, up to a numerical prefactor, the scaling in~\eqref{eq:12-4} was proven analytically by~\cite{rieder1967properties} for stochastic Markovian thermostats, and has been generalized for generic thermostats, in presence of an onsite potential and higher dimensions~\cite{nakazawa1968energy,wang2006nonequilibrium,yamamoto2006nonequilibrium,dhar2006heat,dhar2008heat,dhar2016heat}.
In other terms, in an harmonic chain, the energy current $J_e$ is independent of $L$.

\subsubsection{The diffusive case: Fourier's law}

The other classical case is when the energy carriers are scattering with each other, assuming that the mean interaction time separating two successive scattering events is a function of wavenumber, denoted by $\tau_k$. This is the time in which a phonon covers ballistically the distance $l_k=v_k \tau_k$, the so-called mean free path.
In this case,
 the expression of $J_e$ and $\kappa_e$~\eqref{eq:J12}-\eqref{eq:12-3} are locally defined at a given point $x$. 
 We describe this case following the arguments given in~\cite{feynman2011feynman}, where $e_k^+$ represents the density of phonons to the left of point $x$ that will travel past $x$ in one ballistic flight, i.e., $e_k^+(x-\Delta x)$. Likewise, $e_k^-$ is interpreted as the density to the right of $x$, i.e., $e_k^-(x+\Delta x)$. Assuming that $e^+_k(x)=e^-_k(x)=e_k(x)=c_L T(x)/(2\pi)$ (local equipartition of left- and right-going modes) and that $T(x)$ is a smooth function of $x$, we can expand to first order to obtain
\begin{equation}\label{eq:12-5}
\begin{aligned}
	j_k(x) 
	\simeq -\frac{\partial e_k(x)}{\partial x} 2\Delta x |v_k| 
	\simeq -\frac{\partial T(x)}{\partial x} \frac{c_L }{2\pi} 2l_k |v_k| \,,
\end{aligned}
\end{equation}
where we fix $\Delta x \simeq l_k$ to account for the distance from which the phonons with wavenumber $k$ reach point $x$ within one ballistic flight.
Finally, integrating over $k$ gives
\begin{equation}\label{eq:12-6}
\begin{aligned}
	J_e(x) = \int_{-\pi}^\pi j_k(x) \,dk 
		=- \kappa_e \,\frac{\partial T(x)}{\partial x} \,,~ \text{with} \;\; \kappa_e \simeq \frac1{2\pi} \int_{-\pi}^\pi 2 l_k c_L |v_k|\,dk \,,
\end{aligned}
\end{equation}
which is known as Fourier's law of thermal conduction~\cite{baron1822theorie}: the rate of heat transfer through a material is proportional to the negative gradient of the temperature; the constant of proportionality is the thermal conductivity $\kappa_e$.
Now, the continuity equation for the energy density yields
\begin{equation}\label{eq:12-7}
\begin{aligned}
	&\frac{\partial}{\partial t} \int_{-\pi}^\pi e_k(x) dk = - \frac{\partial J_e(x)}{\partial x}\\
	&\frac{\partial T(x)}{\partial t}   = D \,\frac{\partial^2 T(x)}{\partial x^2}  \,,~\text{with}~ D=\frac{\kappa_e}{c_L}\,.
\end{aligned}
\end{equation}
The diffusion coefficient of the resulting heat equation is given by the thermal conductivity itself, up to a dimensional factor.

Eqs.~\eqref{eq:12-4} and~\eqref{eq:12-6} are heuristic, yet insightful. If $l_k$ is replaced by $L$ in~\eqref{eq:12-6}, one retrieves~\eqref{eq:12-4} (neglecting a factor of $2$): indeed, if the mean free path equals the chain length itself, it means that there are no interactions in the bulk just like in the harmonic system. Also notice that both expressions of $\kappa_e$, ballistic and diffusive, are independent of $\Delta T$.

\subsubsection{Anomalous transport}

If for the ballistic transport $\kappa_e\propto L^1$ and for the diffusive transport $\kappa_e\propto L^0$, there is a third kind of thermal transport, termed anomalous~\cite{dhar2008heat,wang2020thermal,benenti2020anomalous,lepri2016thermal}. 
The following observations are some of the key signatures of anomalous behavior with respect to the diffusive picture of Fourier's law~\cite{lepri2016thermal,iubini2020nonequilibrium}:
(i) the temperature profiles show abrupt boundary jumps in proximity of the thermostats. Moreover, the profiles depart from the expected linear profile;
(ii) for large size $L$, the thermal conductivity $\kappa_e$ depends on the system size with a scaling $\kappa_e\propto L^{-\alpha}$  with $0<\alpha<1$;
(iii) the equilibrium current-current correlator decays in time with non-integrable power-law behavior:
\begin{equation}
	\langle J_e(t)J_e(0)\rangle \propto t^{-(1-\delta)}\,, \quad \text{with }0\leq\delta<1\,;
\end{equation}
(iv) the spread of a spatially localized energy perturbation is superdiffusive. The spatial variance of the perturbation grows as
\begin{equation}
	\sigma^2(t)\propto t^\xi\,, \quad \text{with }\xi>1\,.
\end{equation}
The three coefficients quantify different aspects of the same phenomenon. By different means~\cite{lepri2016thermal}, one can show that they relate to each other via the simple expressions
\begin{equation}
	\delta = \alpha\,,\qquad \xi = 1+\alpha\,.
\end{equation}

A simple way to understand how this behavior may arise, following e.g.~\cite{dhar2008heat}, is to assume that $\tau_k\propto k^{-a}$, with $a>1$. Then, we have that $v_k l_k\propto k^{-a}$, and all the modes for which $k<CL^{-1/a}$ (for a constant $C>0$) are surely ballistic, since with $l_k>L$  they cross on average the whole chain without ever interacting. Thus, we can decompose the conductivity as
\begin{equation}\label{eq:12-7}
	\kappa_e \simeq \frac{L}{2\pi}\int _{-CL^{-1/a}}^{CL^{-1/a}} c_L |v_k| \,dk + \frac{1}{2\pi}\int _{|k|>CL^{-1/a}} 2l_k c_L |v_k| \,dk,.
\end{equation}
Even assuming that all the modes $|k|>CL^{-1/a}$ are diffusive, this gives:
\begin{equation}\label{eq:12-7}
	\kappa_e \simeq L C_1 L^{-\frac{1}{a}}  + C_2  =  C_1 L^{\frac{a-1}{a}}+ C_2\stackrel{L\to\infty}{\sim}{C_1 L^{\frac{a-1}{a}}}\,,
\end{equation}
where $C_1$ and $C_2$ are positive constants. This shows anomalous conduction behavior with an exponent $\alpha=\tfrac{a-1}{a}$, that depends directly on the scaling of $\tau_k$. 


Despite being heuristic, the reasoning above highligths the need for a theory that is able to connect the microscopic properties of the chain with the phononic interaction properties, e.g., $\tau_k$ and $l_k$. Then, we can hope to express $\kappa_e$ directly as a function of the microscopic properties of the system.
For such an endeavor, it is natural to consider the wave kinetic equation for a crystal lattice.
Rudolf Peierls was the first to envision this project for 3D lattices and to consider an evolution equation for the evolution of the statistics of the Fourier amplitudes~\cite{peierls}.
The Peierls-Boltzmann theory is to this day one of the pillars of thermal conduction in 3D solids~\cite{ziman2001electrons,ziman1972principles}.


\subsection{Wave kinetic theory of thermal transport in $\beta$-FPUT chains}

\subsubsection{Reduction of the collision integral to a single integration}

We consider the wave kinetic equation associated to the $\beta$-FPUT model, equation ~\eqref{eq:contwke}, formally analogous to the Peierls-Boltzmann equation. For theoretical and numerical purposes, the multidimensional collision integral, can be reduced to a single integral ~\cite{pereverzev2003fermi,lukkarinen2016kinetic,lukkarinen2008anomalous}.
Indeed, the constraint imposed by integration on the resonant manifold reduces the triple integral of \eqref{eq:contwke} to a one-dimensional integral. There are three types of solution:
 (i) $k_1 = k_3$, $k_2=k_4$\,; (ii) $k_1 = k_4$, $k_2=k_3$\,; 
 (iii) $k_2=h(k_1,k_3) \mod 2\pi$, where
\begin{equation}\label{eq:h}
	h(k_1,k_3) = \frac{k_3-k_1}{2} + 2 \arcsin\left( \tan{\frac{|k_3-k_1|}{4}} \cos \frac{k_3+k_1}{4} \right)\,.
\end{equation}
The first two types  correspond to trivial resonances that contribute to the nonlinear frequency shift and to the nonlinear broadening~\cite{lvov2018double}. The third type of solutions is named non-perturbative and represents non-trivial resonances that are responsible for irreversible spectral transfers.
Pereverzev~\cite{pereverzev2003fermi} noted that the non-perturbative solutions are Umklapp resonances, i.e. such that $k_1+k_2=k_3+k_4+2\pi n$, with $n$ a nonzero integer number. The normal resonances with $n=0$, instead, all result into trivial processes involving only two wavenumbers, with no associated spectral transfers.
By integrating analytically in $k_1$ and $k_3$, integrating out the two delta functions, the following non-vanishing contribution is obtained~\cite{lukkarinen2008anomalous}
\begin{equation}\label{eq:8m}
\begin{aligned}
\mathcal I_k&= \int_0^{2\pi} dk_2 \int_0^{2\pi}dk_1 \; g(k,k_1,k_2)\frac{\delta(k_1-h(k,k_2))}{|\partial_{k_1}\Omega(k,h(k,k_2),k_2)|}\\
&= \int_0^{2\pi} dk_2 \; \frac{g(k,h(k,k_2),k_2)}{\sqrt{\left(\cos\tfrac{k}{2}+\cos\tfrac{k_2}{2}\right)^2 + 4 \sin\tfrac{k}{2}\sin\tfrac{k_2}{2}}}\,,\qquad \text{where}\\
g(k,k_1,k_2) & = |T_{k,k_1,k_2,k+k_1-k_2}|^2 n_kn_{1}n_{2}n_{k+k_1-k_2}
				\left(\frac{1}{n_k}+\frac{1}{n_{1}}-\frac{1}{n_{2}}-\frac{1}{n_{k+k_1-k_2}} \right).
\end{aligned}
\end{equation}

\subsubsection{Small equilibrium perturbation and the linearized collision operator}

We have seen in Sec.~\ref{sec:2} that the general equilibrium state for Eq.~\eqref{eq:contwke} is given by $n_k=T/(\omega_k+\mu)$. Here, we consider $\mu=0$ for simplicity. Then, we consider a small homogeneous perturbation of equilibrium of the form
\begin{equation}\label{eq:eqpert}
	n_k(t) = \bar n_k + \epsilon \tilde n_k\,,\qquad \text{with}\quad \bar n_k = \frac{T}{\omega_k}\,.
\end{equation}
Plugging this into~\eqref{eq:contwke}, one obtains ${\partial \tilde n_k}/{\partial t} = \mathcal L \tilde n_k\,$, with
    \begin{equation}
	\mathcal{L}f_k = \frac{9}{4}\pi T^2\int_{0}^{2\pi} \left(\omega_3^2 f_{3} + \omega_2^2 f_{2} - \omega_1^2 f_{1} - \omega_k^2 f_{k}\right)\delta(\Delta K)\delta(\Delta \Omega)dk_1dk_2dk_3\,,
  \label{eqn:lin}
\end{equation}
where $\mathcal L$ is the linearized collision operator ~\cite{lukkarinen2016kinetic,mellet2015anomalous}.
Now, by the same technique employed to derive Eq.~\eqref{eq:8m},  the linearized collision integral can be reduced to a single integration, which reads
\begin{equation}
  \label{eqn:lin-sing}
  \begin{aligned}
    \mathcal{L}f_k = \frac{9}{4}\pi T^2\int_{0}^{2\pi} \frac{\omega_{k+h(k,k')-k'}^2 f_{k+h(k,k')-k'} + \omega_{k'}^2 f_{k'} - \omega_{h(k,k')}^2 f_{h(k,k')} - \omega_k^2 f_{k}}{\sqrt{\left(\cos\tfrac{k}{2}+\cos\tfrac{k'}{2}\right)^2 + 4 \sin\tfrac{k}{2}\sin\tfrac{k'}{2}}} dk'\,,
  \end{aligned}
\end{equation}
where $h(k,k')$ is given by Eq.~\eqref{eq:h}.

\subsubsection{Relaxation-time approximation}

As seen in Section \ref{sec:WT}, the wave kinetic equation~\eqref{eq:contwke} is decomposed into an ``integral part'' $\eta_k$, which
includes all the terms of the collision integral that do not contain $n_k$, and a ``multiplicative part'' $-\gamma_k n_k$. The interpretation of this decomposition is simple: the integral part is positive and corresponds to ``creation'' of waves of wavenumber $k$ due to quartets in which $k$ is one of the products of the scattering, hence a positive net change of wave-action $n_k$ per unit of time; the multiplicative part is negative and corresponds to ``absorption'' of wavenumber $k$ in the scattering quartet, hence implying a negative net change of wave-action. 

Analogously, after considering a small perturbation of the equilibrium state, one can proceed to decompose the linearized wave kinetic equation~\eqref{eqn:lin} as
\begin{equation}\label{eq:apprt}
	\frac{\partial \tilde n_k}{\partial t} = \tilde \eta_k - \tilde\gamma_k \tilde n_k\,,
\end{equation}
where $\tilde \eta_k$ and $-\tilde\gamma_k \tilde n_k$ are the integral and the multiplicative parts, respectively.
Now, if it is supposed that the initial perturbation is narrowly localized around wavenumber $k$, i.e., $\tilde n_k\simeq0$ elsewhere, we shall approximate $\tilde \eta_k\simeq0$ (since the integrand is vanishing almost everywhere) and obtain
\begin{equation}\label{eq:gamma}
	\tilde \gamma_k =  {9}\pi T^2 \sin^2\tfrac{k}{2}\int_{0}^{2\pi} \frac{dk'}{\sqrt{\left(\cos\tfrac{k}{2}+\cos\tfrac{k'}{2}\right)^2 + 4 \sin\tfrac{k}{2}\sin\tfrac{k'}{2}}} \,.
\end{equation}
This is the so-called {\it relaxation-time} approximation to equation~\eqref{eq:apprt}, with solution given by
\begin{equation}
	\tilde n_k(t) = \tilde n_k(0) e^{-t/\tau_k}\,, \qquad \text{with }\quad \tau_k=\tilde\gamma_k^{-1}\,.
\end{equation}

The relaxation-time approximation has been widely used for heat conduction in insulators~\cite{herring1954role,klemens1956thermal,beck1975dynamical}. Pereverzev~\cite{pereverzev2003fermi} argues that this approximation describes a physical situation in which all modes except mode $k$ are in equilibrium. In general this is not the case. However, the evolution of the small wavenumbers is very slow. Therefore, for any initial nonequilibrium distribution all modes except those with small $k$ quickly relax to equilibrium. As a result, for the small-$k$ modes, after a short time, the physical situation is similar to the one described by the relaxation-time approximation, and the integral part of the collision operator becomes negligible compared to the multiplicative part. Following this plausibility argument, it is possible to compute the relaxation time $\tau_k$, in the limit $k\to0 \mod2\pi$.
 Starting from~\eqref{eq:gamma}, after some non-trivial algebra~\cite{pereverzev2003fermi,nickel2007solution,lukkarinen2008anomalous}, one finally obtains
\begin{equation}
	\tau_k\propto k^{-5/3}\,.
\end{equation}
It has to be mentioned that an analogous result for the relaxation time had been previously obtained also via Mode Coupling Theory (MCT)~\cite{pomeau1975time,kubo2012statistical,ernst1991mode,lepri1998relaxation,lepri2003thermal,delfini2006self},  first introduced to deal with long-time tails in fluids~\cite{pomeau1975time}, dictated by the slow time scales of the small wavenumbers. MCT consists of deriving effective generalized Langevin equations describing the ``diffusion'' of macroscopically conserved quantities. The memory terms in these equations are estimated via Mori-Zwanzig projection operators~\cite{kubo2012statistical} and self-consistent approximations. 
Given the attempt to estimate the same quantity with emphasis on the small wavenumbers, it is not a surprize that the independent results from MCT and the wave kinetic equation are in agreement. In addition, the scaling of $\tau_k$ has also solid numerical verification~\cite{lepri2016thermal}.

\subsubsection{A rigorous formula for $\kappa_e$: the Green-Kubo formula}

At this point, regardless of how $\tau_k$ has been estimated, the common way to use this information to obtain a scaling behavior for the conductivity is to appeal to the {\it Linear Response Theory}~\cite{lepri2003thermal,pereverzev2003fermi,lukkarinen2008anomalous,dhar2008heat,marconi2008fluctuation}. In this framework, the goal is to obtain the ``transport coefficient'' that quantifies the response of the system to a perturbation from an external field. 
The transport coefficient is typically calculated via the so-called Green-Kubo formula~\cite{kubo2012statistical}, which relates the nonequilibrium response to time correlations in the equilibrium state. 
Assuming small temperature gradients, the following Green-Kubo formula can be derived:
\begin{equation}\label{eq:GK}
	\kappa_e = \frac{1}{c_L T^2}\lim_{\tau\to\infty}\lim_{L\to\infty}\int_0^\tau dt C(t)\,.
\end{equation}
$C(t)=\langle \mathcal J(0) \mathcal J(t)\rangle $ is the current-current time correlation; the total current is given by $\mathcal J(t)=\int_0^L dx J_e(x,t)$, where also time dependence is now allowed. 
$\langle\cdot\rangle$ denotes ensamble averaging~\cite{dhar2008heat,lukkarinen2016kinetic}. 
It was shown in~\cite{pereverzev2003fermi} that in the relaxation-time approximation the time correlation function reads
\begin{equation}\label{eq:rta}
	C(t) = \frac{2c_L^2 T^2}{\pi}\int_0^\pi dk e^{-\frac{t}{\tau_k}}v_k^2 \propto t^{-3/5}\,.
\end{equation}
Now, an upper cutoff is imposed in the time integration in~\eqref{eq:GK}, $\tau\propto L$, invoking the finite speed of propagation and a finite system size $L$~\cite{lepri2003thermal, pereverzev2003fermi,nickel2007solution,lukkarinen2008anomalous}. As a result, we obtain
\begin{equation}\label{eq:scaling25}
	\kappa_e \propto L^{2/5}\,,
\end{equation}
consistent with the value typically obtained from the direct numerical simulations~\cite{lepri2003universality,wang2011power,dematteis2020coexistence}. In ~\cite{dhar2008heat} it has been noted that the same result could be obtained by setting $a=5/3$ in the simple-minded calculation~\eqref{eq:12-7}, where an analogous cutoff was used. 

\subsubsection{Toward the macroscopic equations: homogeneous perturbation}

A step forward in the formalization of the thermal conduction from the collision integral is taken in~\cite{aoki2006energy,lukkarinen2008anomalous,lukkarinen2016kinetic}. In these works, a generalization of the $\beta-$FPUT model is considered to take into account the presence of an onsite potential with relative ``strength'' $\delta$, which takes values in the interval $(0,1/2]$. The generalized dispersion relation is given by
\begin{equation}
	\omega(k)=(2-4\delta\cos k)^{1/2}.
\end{equation}
When $\delta=1/2$, the dispersion relation reduces to the $\beta-$FPUT case: $\omega(k)=2\sin (k/2)$. 
An analytical generalization of Eq.~\eqref{eq:8m} has been obtained~\cite{aoki2006energy}, in which the integrand denominator is replaced by $\delta\cdot(\omega(k_1)^{-1}\sin k_1 - \omega(k+k_1-k_2)^{-1}\sin (k+k_1-k_2))$, for $k_1=h(k,k_2)$ ad in which $h(x,y)$ has a different expression that reduces to Eq.~\eqref{eq:h} for $\delta=1/2$. 

Here, we summarize the main steps of the derivations in~\cite{aoki2006energy,lukkarinen2008anomalous,lukkarinen2016kinetic}. The starting point is again the Green-Kubo formula~\eqref{eq:GK}. Because the thermodynamic limit has already been taken to obtain the wave kinetic equation in section~\ref{sec:WT}, Eq. \eqref{eq:GK} reads as:
\begin{equation}\label{eq:GK2}
	\kappa_e = \frac{1}{c_L T^2}\int_0^\infty dt C(t)\,.
\end{equation}
Neglecting the cubic part of the Hamiltonian, the current-current correlation is shown to take the form
\begin{equation}\label{eq:GK3}
	C(t) = \int_0^{2\pi} dk\; v_k \omega_k \tilde n_k (t)\,,\qquad \tilde n_k(0)=v_k\omega_k \bar n_k^2\,,
\end{equation}
where $\tilde n_k(t)$ is the small spatially homogeneous perturbation~\eqref{eq:eqpert}, but now constrained to have a given form. In particular, this form is essentially proportional to $\omega_k^{-1}$, the mode of propagation corresponding to energy conservation~\cite{lukkarinen2016kinetic,pereverzev2003fermi,mellet2015anomalous}. Formula~\eqref{eq:GK3} is intuitive: the averaged magnitude of the energy perturbation is weighted by its speed of propagation separately for each wavenumber, and then integrated over all wavenumbers. The equation for the time evolution of the perturbation is given by
\begin{equation}
	\frac{\partial \tilde n_k}{\partial t} = \mathcal L \tilde n_k = \mathbb L\bar n^{-2} n_k\,,
\end{equation}
defining a new operator $\mathbb L$ that does not depend on $\omega_k$.
Then, using the notation suggested by the form of the linearized collision operator~\cite{aoki2006energy,lukkarinen2008anomalous}, one defines:
\begin{equation}
	\tilde h_k = \bar n_k^{-1}\tilde n_k\,,\qquad  \tilde{\mathbb  L} = \bar n_k^{-1} \mathbb L \bar n_k^{-1}\,,
\end{equation}
where $\tilde{ \mathbb L}$ is proven to be a positive operator on $L^2([0,2\pi])$.
Now, the renormalized perturbation simply evolves according to
\begin{equation}
	\frac{\partial \tilde h_k }{\partial t}= \tilde {\mathbb  L} \tilde h_k \,,\qquad \tilde h_k(0)=v_k\omega_k \bar n_k\,,
\end{equation}
with solution
\begin{equation}
	\tilde h_k(t) = e^{-t \tilde {\mathbb  L}} \tilde h_k(0)\,.
\end{equation}
Therefore,
\begin{equation}
	\tilde n_k(t) = \bar n_k e^{-t \tilde {\mathbb  L}}\bar n_k^{-1} \tilde n_k(0)\,,
\end{equation}
and by direct substitution in~\eqref{eq:GK3}, we obtain
\begin{equation}\label{eq:GK4}
\begin{aligned}
	C(t) = \int_0^{2\pi} dk\; v_k \omega_k \bar n_k e^{-t \tilde {\mathbb  L}}\bar n_k^{-1} \tilde n_k(0)
	      = \int_0^{2\pi} dk  \langle \tilde h_k(0), e^{-t\tilde {\mathbb  L}} \tilde h_k(0)\rangle\,,
\end{aligned}
\end{equation}
where $\langle \cdot,\cdot \rangle$ denotes the scalar product in $L^2([0,2\pi])$. Substituting in~\eqref{eq:GK2}, we have
\begin{equation}\label{eq:GK5}
\begin{aligned}
	\kappa_e = T^{-2}\int_0^\infty dt\; \langle \tilde h_k(0), e^{-t\tilde {\mathbb  L}} \tilde h_k(0)\rangle
		  =  T^{-2} \langle \tilde h_k(0), \tilde {\mathbb  L}^{-1} \tilde h_k(0)\rangle\,.
\end{aligned}
\end{equation}
which connects elegantly the thermal conductivity to the inverse linearized collision operator, for details see e.g.~\cite{mellet2015anomalous}.
In~\cite{aoki2006energy} it is proven that expression~\eqref{eq:GK5} gives a finite value for $\kappa_e$ in the case of onsite pinning potential. On the other hand, in~\cite{lukkarinen2008anomalous} the conductivity $\kappa_e$ of the $\beta$-FPUT unpinned chain is shown to diverge with the system size because the operator $L$ is vanishing for small wavenumber. Therefore, the authors of~\cite{lukkarinen2008anomalous} appeal to a relaxation time approximation analogous to~\eqref{eq:rta} and obtain a scaling consistent with~\eqref{eq:scaling25}.


\subsubsection{Toward the macroscopic equations: inhomogeneous perturbation}\label{sec:diff}

In the previous sections, the homogeneous case has been considered. In actual experiments, boundaries introduce inhomogeneity.
In inhomogeneous problems, spatial transport should be added in the kinetic equation.
Two alternative phenomenological approaches 
to deal with inhomogeneous one-dimensional heat transport have been considered: (i) surrogate stachastic processes like L\'evy walks~\cite{cipriani2005anomalous,zhao2006identifying,zaburdaev2011perturbation,lepri2011density,liu2014anomalous,kundu2019fractional}; (ii) nonlinear fluctuating hydrodynamics~\cite{spohn2016fluctuating,mendl2013dynamic,das2014numerical,cividini2017temperature} -- see also~\cite{benenti2020anomalous,dhar2019anomalous} for recent reviews on the topic. 
In the first case, phonons are modelled as heat carriers performing ballistic L\'evy flights. The second approach is similar to mode coupling theory, but it additionally includes nonlinear coupling coefficients between the different modes of propagation; the coefficients are derived self-consistently from the fluctuation-dissipation relations. In both cases, the spread of the energy perturbations (or equivalently of space time correlations) is governed by a fractionary diffusion equation with a spatial derivative of order $\nu$, with $1<\nu<2$ ($\nu=2$ for regular diffusion and $\nu=1$ for pure ballistic propagation)~\cite{crouseilles2016numerical}. A superdiffusive behavior with exponent $\nu$ is associated with a conductivity exponent $\alpha$ via $\nu=2-\alpha$~\cite{dhar2019anomalous}. For the $\beta$-FPUT model, both approaches lead to a superdiffusive heat peak with $\nu=5/3$, corresponding to an anomalous conductivity exponent $\alpha=1/3$~\cite{dhar2019anomalous,spohn2014nonlinear}. To date, there is no universally accepted value of $\alpha$  for $\beta$-FPUT, although there is indication for a crossover from $\alpha\simeq2/5$ at low temperatures (in agreement with~\eqref{eq:scaling25}) to $\alpha\simeq1/3$ at high temperatures~\cite{lepri2016thermal}. Independent methods based on a small noise perturbation lead to superdiffusion with $\nu=3/2$~\cite{jara2015superdiffusion,basile2016thermal}.

In~\cite{mellet2015anomalous}, the results of~\cite{lukkarinen2008anomalous} are generalized to spatially inhomogeneous perturbations: a superdiffusive fractionary heat equation is thus derived from the linearized collision operator. This amounts to a rigorous derivation of the macroscopic equations from the mesoscopic wave kinetic picture, in the same vein as~\cite{bonetto2000fourier,saint2009hydrodynamic} for the Boltzmann equation. The resulting fractionary diffusion equation has exponent $\nu=8/5$~\cite{mellet2015anomalous}, corresponding to a Green-Kubo correlation with the same scaling of Eq.~\eqref{eq:rta}, i.e. $C(t)\propto t^{-3/5}$, and to a spread of the variance of localized initial perturbations that goes like $t^{7/5}$. Indeed, in the diffusive case the variance would spread like $t^1$ and in the ballistic case like $t^2$.

\subsubsection{Local spectral analysis from simulations of the microscopic dynamics}\label{sec:loc}
%

The work~\cite{dematteis2020coexistence} has recently proposed a new perspective by performing a spatially local Fourier analysis in direct numerical simulations of a long $\beta$-FPUT chain at low temperature. The Fourier transform is calculated in an $O(\sqrt L)$-wide sliding window, as opposed to the whole chain -- as previously done, e.g., in~\cite{lepri2005studies}. This procedure allows  to analyse the spectral properties along the chain in a {\it local} sense.
\begin{figure} [hbt] 
\centering
\includegraphics[width=0.49\linewidth]{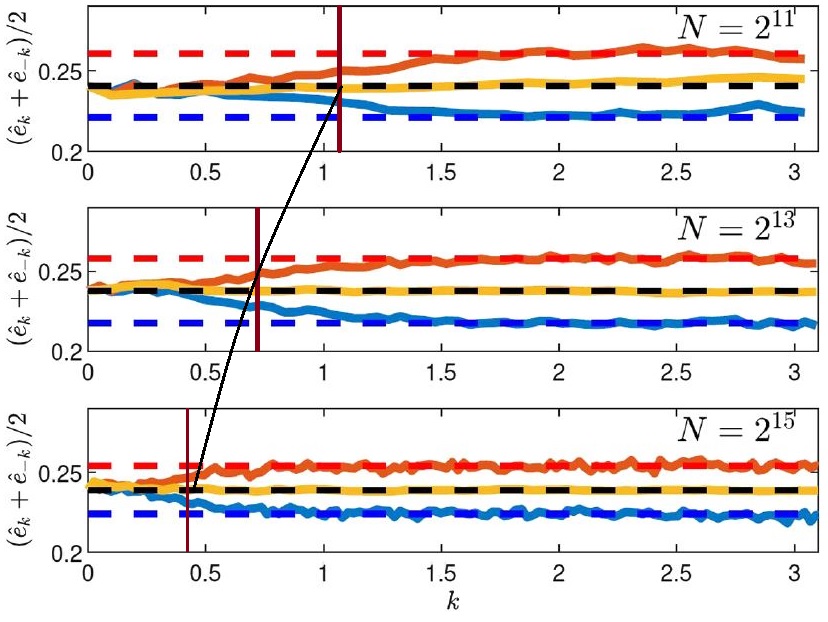}%
\includegraphics[width=0.51\linewidth]{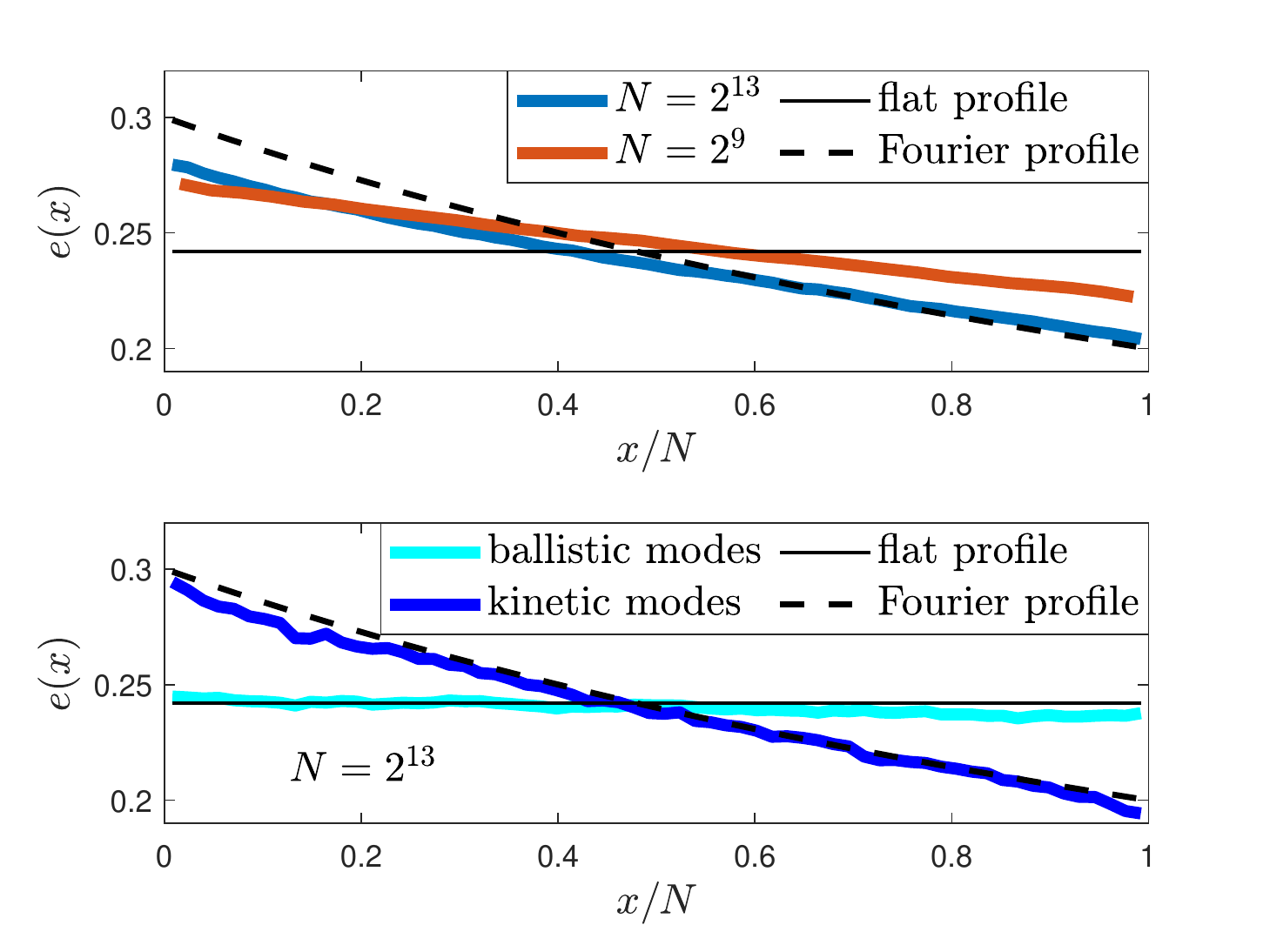}
 \caption{ Left: The solid lines are the symmetrized
stationary local energy spectra computed in a sliding window of width $\sqrt{L}$,
centered at $x_1=0.2L$ (red) and at $x_2=0.8L$ (blue). The
dashed lines with same colors are at the respective average energy
per particle at $x_1$ and $x_2$. The yellow line is the global energy spectrum. 
To guide the eye, the black solid curve follows a scaling proportional to $k^{-3/10}$. 
Upper right: Numerical profiles of $e(x)$ for different $L$. Lower right: the contributions to the blue curve of the upper panel is decomposed into its contributions from
low wavenumbers (light blue) and high wavenumbers (dark blue), normalized by the fraction of modes in each set: $k_c/\pi$ and $(\pi-k_c)/\pi$, respectively. 
Figure  taken from~\cite{dematteis2020coexistence}.}
 \label{fig:local-FT}
\end{figure}
%

The left panel of Fig.~\ref{fig:local-FT} shows the change in the local spectral content around the two opposite ends of the chain, which are thermostatted at temperatures $T_+$ and $T_-$. The high wavenumbers are locally thermalized (in energy equipartition) at the local temperature given by a spatial Fourier profile that interpolates linearly between $T_+$ and $T_-$. On the contrary, the low wavenumbers mantain the same temperature $(\bar T = T_++T_-)/2$ throughout the chain, indicating a lack of interactions. This is confirmed by the observed filtered Fourier profiles in the right panel of Fig.~\ref{fig:local-FT}, in which the high-pass filtered temperature tends to the Fourier profile, whereas the low-pass temperature tends to a flat profile at temperature $\bar T$. These observations are qualitatively consistent with the heuristic decomposition~\eqref{eq:12-7}, in which the low wavenumbers are ballistic and the high wavenumbers are diffusive. 
However, it turns out   that the critical separation between ballistic and diffusive behavior, say $k_c(L)$, is consistent with $k_c\propto L^{-3/10}$, rather than $k_c\propto L^{-3/5}$. 
In~\cite{dematteis2020coexistence}, 
an interpretation of this scaling is given such that 
ballistic and diffusive modes are  differentiated upon based on their local relaxation properties.

The spectral analysis of~\cite{dematteis2020coexistence} also extends to the energy flux. 
In particular, the low wavenumbers preserve their ballistic flux content as in the pure harmonic chain, while for the high wavenumbers the flux content decreases in inverse proportion to $L$ according to the Fourier's law.

The explanation of the scaling of $\kappa_e(L)$ given in~\cite{dematteis2020coexistence} includes a non-constant transmission coefficient between the thermostat and the chain that is linear in $k$ (instead of a constant $e_k^+-e_k^-$ as in Eq.~\eqref{eq:12-7}), returning the expected anomalous exponent $\alpha=2/5$.


\subsubsection{Numerical simulation of the inhomogeneous wave kinetic equation}
The results reviewed in Section~\ref{sec:loc} were corroborated by a direct numerical simulation of the spatially inhomogeneous wave kinetic equation associated with the $\beta$-FPUT system~\cite{de2022anomalous}. 
The wave kinetic equation in the inhomogeneous case reads:
\begin{equation}
	\frac{\partial  n_k}{\partial t} +v_k\frac{\partial  n_k}{\partial x}= 4\pi
\int_{0}^{2\pi} |T_{k,1,2,3}|^2 n_kn_1n_2n_3\left(\frac{1}{n_k}+\frac{1}{n_1}-\frac{1}{n_2}-\frac{1}{n_3}\right)
\delta_{k,1}^{2,3}\delta({\omega_{k,1}^{2,3}}) dk_{123}
\label{eq:nhwke}
\end{equation}
A rigorous justification of the transport term can be found in \cite{ampatzoglou2021derivation} for the Nonlinear Schr\"odinger equation.
\begin{figure}[hbt] 
\centering
\includegraphics[width=0.75\linewidth]{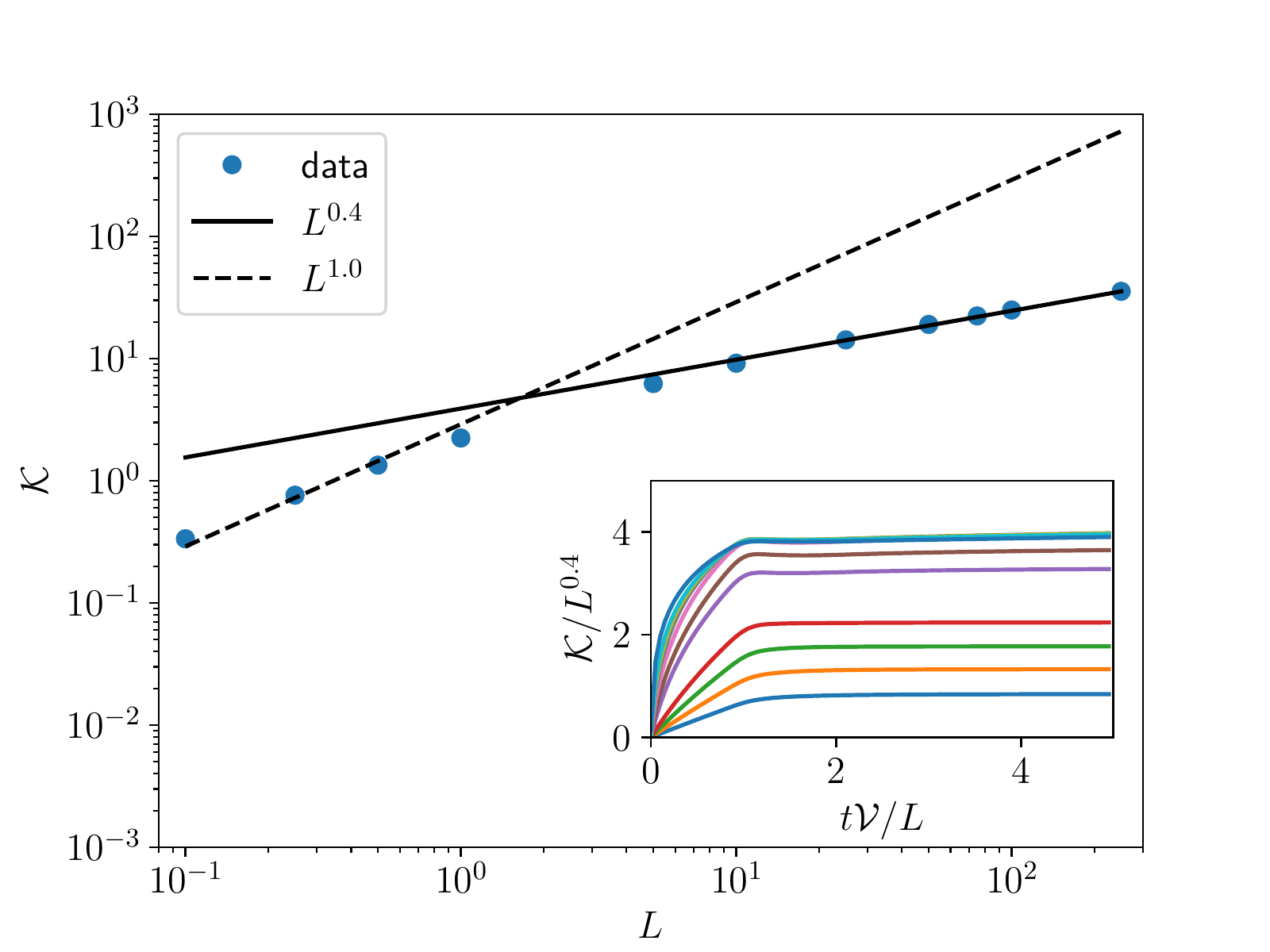}%
 \caption{Results from numerical simulation of the inhomogeneous wave kinetic equation of $\beta$-FPUT taken from~\cite{de2022anomalous}. 
Shown is the size-dependent scaling of $\kappa_e$ converging to the $L^1$ scaling for small $L$ and to the $L^{2/5}$ scaling for large $L$. The inset shows the time evolution of the conductivity for the different values of $L$.
}
 \label{fig:FT-flux2}
\end{figure}
The energy flux spectral content was shown to be globally consistent with the results from the microscopic dynamics in~\cite{dematteis2020coexistence}.
Numerical results of equation (\ref{eq:nhwke}) presented show that the steady-state conductivity is shown to go from a $L^\alpha$ scaling with $\alpha=1$  at small $L$, when the size is too small for nonlinearity to play a role and the system is essentially linear, to a $L^{\alpha}$ with $\alpha=2/5$ scaling at large $L$, see Fig.~\ref{fig:FT-flux2}. Under the approximation that $\hat j(k)\propto k$ as $k\to0$, the $\alpha=2/5$ scaling is consistent with $k_c\propto L^{-3/10}$, as already mentioned~\cite{dematteis2020coexistence}. 

The fractional diffusion picture reviewed in Section \ref{sec:diff} focuses on the superdiffusive character of the central heat peak to describe the macroscopic effect of anomalous conduction, finding that the contribution to the energy transport by the acoustic peaks is asymptotically negligible~\cite{lepri2005studies}.

A slightly different picture emerges from the Wave Turbulence framework of~\cite{dematteis2020coexistence,de2022anomalous}. In the proposed two-state ballistic-diffusive simplifying interpretation, the divergent contribution of the thermal conductivity is due to the modes in the ballistic peaks, while the heat peak is viewed as roughly diffusive. Notice that a diffusive heat peak does not seem inconsistent from the scaling predicted by nonlinear fluctuating hydrodynamics, see e.g. Eq.~(4.16) in~\cite{spohn2016fluctuating}. In this picture, the observed macroscopic scaling $\alpha=2/5$ is retrieved because of the dependence of the separation between the two sets of modes on the system size, $k_c(L)$. Interestingly, this interpretation shifts the focus of the anomalous energy transport from the heat peak to the acoustic peaks, in a way that: (i) remains consistent with the observed anomalous scaling of the conductivity; (ii) suggests a potential analogy with the phenomenon of second sound~\cite{chester1963second,chandrasekharaiah1986thermoelasticity,prohofsky1964second,huberman2019observation}.
 By means of response theory, the works~\cite{kuzkin2020ballistic,fernando2020non,kuzkin2021unsteady,bohm2022analysis} have recently proposed an independent approach to thermal transport in 1D particle chains, with a focus on the coherent persistence of the ballistic excitations. More studies are needed to settle these questions.

%
%
%
%
%
%
%


\section{Conclusions}
In this report, we have given an overview of the current state of
the art with respect to the application of Wave Turbulence theory to one-dimensional particle chains.  Most of the results reported are related to  the Fermi-Pasta-Ulam-Tsingou model, both $\alpha$ (quadratic
nonlinearity) and $\beta$ (cubic nonlinearity); nontheless, other models like the Nonlinear
Discrete Klein Gordon have been discussed in the framework of the Wave Turbulence approach.
Nonlinear one-dimensional chains are key models for condensed-matter studies, and they have been (and in fact they are) of paramount importance in the development of statistical and nonlinear physics.
We have reviewed how diverse fields like Hamiltonian chaos, solitons and computational analysis have flourished  in relation with these systems.
In parallel with these developments, the statistical theory of Hamiltonian systems composed by weakly interacting waves has been developed, that is the Wave Turbulence approach. 

Here, we have reviewed how the direct application of Wave Turbulence to the dynamics of one-dimensional chains may give important physical insights, when the nonlinearity is weak and a perturbative approach is meaningful. 
In particular, we have shown that Wave Turbulence allows for the understanding of the mechanisms underlying the thermalization of nonlinear chains, as well as to predict the power-law scalings of the thermalisation-time encountered in numerical simulations. It is underlined that  resonances are the main ingredient of the dynamics.
Furthermore, in Wave Turbulence theory it is natural to find and quantify subtle features such as the shift and broadening in frequency due to nonlinearity, which allows to predict the Chirikov overlap frequency leading to the predominancy of quasi-resonances.
In a similar vein, we have reviewed the recent attempts to shed some new light on the problem of anomalous thermal conduction via Wave Turbulence.

The story of Wave Turbulence in application to these systems is far
from being complete. Here are some of the questions that might deserve
further consideration:

\begin{itemize}
\item The mathematical research on the wave kinetic equation has received enormous attention in the last years~\cite{staffilani2021wave,lukkarinen2011weakly,buckmaster2021onset,dymov2021formal,deng2021derivation}. 
Exact mathematical results are of paramount importance for establishing the regime at which 
Wave Turbulence is expected to work. Nowadays, there is no rigorous derivation of the Wave Kinetic equation
for the models described in this review. Moreover, as far as we know, proofs of convergence of the near identity
canonical transformation needed to remove non resonant terms (and write the dynamical equation in a form suitable for the applicability of the Wave Turbulence theory) have not been accomplished. 
\item In the pioneering work by Peierls~\cite{peierls1929}, the N-mode PDF of the field was at the core of the description of the statistical behaviour of waves, as in the important work by Zaslavskii and Sagdeev~\cite{zs}.
In the following, the research in the field has focused on the spectrum, a more practical quantity to analyze.
Yet, in the last years a renewed interest in the PDF has grown, in order to develop a more rigorous approach, and also to gain new physical insights on the role of fluctuations~\cite{choi2005anomalous,choi2005joint,eyink2012kinetic,nazarenko2011wave,chibbaro2017wave,chibbaro20184,rosenzweig2022uniqueness}.
Research in this direction will be certainly fruitful~\cite{deng2021propagation,guioth2022path,hrabski2022properties}, and one-dimensional chains are the ideal playground.

\item An important issue, not considered in this report, is the presence of forcing and damping in
  these discrete chains. A detailed numerical study of such settings is
   yet to come. Is there any universality in such forced and
  damped spectra? Are there any power-laws? Are there any Zakharov-Kolmogorov
  spectra? In principle, because for chains the dispersion relation is not an homogeneous function, then the formal derivation of the Zakharov-Kolmogorov solutions should not be possible; however, 
  it would be interesting to see if the wave kinetic equation in the presence of forcing and dissipation 
  admits stationary solutions. 
 Similarly, many issues arise naturally when finite-size effects are considered, far from the safer asymptotic limits.  
 
 \item A key feature in condensed matter physics is the presence of disorder~\cite{lieb2013mathematical,chaikin1995principles}, which may lead to localisation in the motion of particles, as in Anderson localisation~\cite{anderson1958random,ishii1973localization}.
 Then, one would like to understand the transport properties of a disordered anharmonic chrystal, that is a one-dimensional chain with some random potential and/or random mass distribution.
Important results on this question have been obtained at a rigorous level~\cite{frohlich1983absence,benettin1988nekhoroshev}, and various attempts to address the issue of the interaction between weak nonlinearity and disorder in such chains have been made~\cite{campbell2004localizing,flach2009universal,ivanchenko2011anderson,mulansky2011strong,fishman2012nonlinear}.
A conventional idea is that sufficient disorder is able to localise the dynamics and transport is anomalously low. Yet, many questions remain unsettled in the case of many particles, and the Wave Turbulence approach has just started to be used~\cite{zhang2017dynamical,nazarenko2019wave,sun2020effects,wang2020wave}.
A systematic analysis of this  problem would be of clear interest, and possible applications are fascinating~\cite{segev2013anderson,guasoni2017incoherent}.
 
\item Although many numerical simulations have been performed with the $\alpha$-, $\beta$-FPUT, the results in the thermodynamic limit, especially for the $\alpha$ are not in excellent agreement with expectations. This result is in agreement with the one discussed in \cite{fu2019universal} where it is mentioned that the Wave Turbulence works better for even potentials. Whether this result is related to finite size effect is not understood yet, and definitely larger simulations should be considered.

\item The validity of the six-wave kinetic equation is based on a subtle balance between nonlinearity and number of particles, and now is basically written as a conjecture. At least a formal derivation of such an equation should be attempted. 

\item For small number of particles,  there are some rigorous proofs of the applicability of the  KAM and Nekhoroshev theorems \cite{henrici2008results}. As far as we know, there is no strong numerical evidence that the equipartition time scale diverges exponentially with the nonlinearity parameter. It would be of paramount importance to establish in terms of $\epsilon$ and $N$ the transition between a power law predicted by the Wave Turbulence theory and the exponential prediction based on Nekhoroshev theorem. 

\item The study of thermalization properties has been mostly limited to one-dimensional chains. There are few exceptions, see for example 
\cite{benettin2005time}. Wave Turbulence has never been applied to two- or three-dimensional chains. We expect that resonances are much more dense with respect to the one-dimensional case and probably there should be no need to look at the six-wave interactions to study the thermalization properties.

\item Concerning the nonhomogeneous problem with two thermostats, the use of the wave kinetic equation is rather new, and numerical computations are at their primordial stage. A direct comparison between the molecular dynamics and the kinetic equation is still lacking; boundary conditions, mimicking the thermostats, deserve also more attention in the comparison between microscopic and macroscopic dynamics. 

\end{itemize}

This list is definitely not exhaustive, and questions of this type
will keep researches busy in the decades to come.

\label{sec:14}

\appendix

\section*{Acknowledgments}
M.O. was funded by Progetti di
Ricerca di Interesse Nazionale (PRIN), Project No. 2020X4T57A, by the European Commission H2020 FET Open
Boheme, grant no. 863179, and by the Simons Foundation, Award 652354 on Wave Turbulence. The authors thank S. Lepri and A. Vulpiani for insightful discussions. 

\bibliographystyle{elsarticle-num} 

\bibliography{bib}





\end{document}